\documentclass{article}

\usepackage{amsfonts}
\usepackage{amsmath}
\usepackage{amssymb}
\usepackage{mathrsfs}
\newtheorem{theorem}{Theorem}

\title{The Structure and Interpretation of Cosmology: Part I - General Relativistic Cosmology}
\author{Gordon McCabe}

\def\eqalign#1{\,\vcenter{\openup.7ex\mathsurround=0pt
 \ialign{\strut\hfil$\displaystyle{##}$&$\displaystyle{{}##}$\hfil
 \crcr#1\crcr}}\,}

\begin{document}

\maketitle

\begin{abstract}

The purpose of this work is to review, clarify, and critically
analyse modern mathematical cosmology. The emphasis is upon
mathematical objects and structures, rather than numerical
computations. This paper concentrates on general relativistic
cosmology. The opening section reviews and clarifies the
Friedmann-Robertson-Walker models of general relativistic
cosmology, while Section 2 deals with the spatially homogeneous
models. Particular attention is paid in these opening sections to
the topological and geometrical aspects of cosmological models.
Section 3 explains how the mathematical formalism can be linked
with astronomical observation. In particular, the informal,
observational notion of the celestial sphere is given a rigorous
mathematical implementation. Part II of this work will concentrate
on inflationary cosmology and quantum cosmology.

\end{abstract}

\section{The Friedmann-Robertson-Walker models}

\subsection{Geometry and topology}

Let us review and clarify the topological and geometrical aspects
of the Friedmann-Robertson-Walker (FRW) models of general
relativistic cosmology. Whilst doing so will contribute to the
overall intention of this paper to clarify, by means of precise
mathematical concepts, the notions of modern cosmology, there are
further philosophical motivations: firstly, to emphasise the
immense variety of possible topologies and geometries for our
universe, consistent with empirical (i.e. astronomical) data; and
secondly, to emphasise the great variety of possible other
universes.

The general interpretational doctrine adopted in this paper can be
referred to as `structuralism', in the sense advocated by Patrick
Suppes (1969), Joseph Sneed (1971), Frederick Suppe (1989), and
others\index{structuralism}. This doctrine asserts that, in
mathematical physics at least, the physical domain of a theory is
conceived to be an instance of a mathematical structure or
collection of mathematical structures. The natural extension of
this principle proposes that an entire physical universe is an
instance of a mathematical structure or collection of mathematical
structures.

Those expressions of structuralism which state that `the' physical
universe is an instance of a mathematical structure, tacitly
assume that our physical universe is the only physical universe.
If one removes this assumption, then structuralism can be taken as
the two-fold claim that (i) our physical universe is an instance
of a mathematical structure, and (ii), other physical universes,
if they exist, are either different instances of the same
mathematical structure, or instances of different mathematical
structures. If some aspects of our physical universe appear to be
contingent, that may indicate how other physical universes provide
different instances of the same mathematical structure possessed
by our universe. Alternatively, given that mathematical structures
are arranged in tree-like hierarchies, other physical universes
may be instances of mathematical structures which are sibling to
the structure possessed by our universe. In other words, the
mathematical structures possessed by other physical universes may
all share a common parent structure, from which they are derived
by virtue of satisfying additional conditions. This would enable
us to infer the mathematical structure of other physical universes
by first generalizing from the mathematical structure of our own,
and then classifying all the possible specializations of the
common, generic structure.

Hence, it is the aim of this paper not only to define the
mathematical structures used in modern cosmology to represent our
universe on large-scales, but to explore the variety of possible
instances of those structures, and to emphasise how those
mathematical structures are special cases of more general
mathematical structures. The intention is to establish the
mathematical structure possessed by our own universe, and to use
that to imply the nature of other universes.

\hfill \break

Geometrically, a FRW model is a $4$-dimensional Lorentzian
manifold $\mathcal{M}$ which can be expressed as a warped product,
(O'Neill 1983, Chapter 12; Heller 1992, Chapter 6):

$$
I \times_R \Sigma \,.
$$ $I$ is an open interval of the pseudo-Euclidean manifold
$\mathbb{R}^{1,1}$, and $\Sigma$ is a complete and connected
$3$-dimensional Riemannian manifold. The warping function $R$ is a
smooth, real-valued, non-negative function upon the open interval
$I$. It will otherwise be known as the scale factor.

If we deonote by $t$ the natural coordinate function upon $I$, and
if we denote the metric tensor on $\Sigma$ as $\gamma$, then the
Lorentzian metric $g$ on $\mathcal{M}$ can be written as

$$
g = -dt \otimes dt + R(t)^2 \gamma \,.
$$

One can consider the open interval $I$ to be the time axis of the
warped product cosmology. The $3$-dimensional manifold $\Sigma$
represents the spatial universe, and the scale factor $R(t)$
determines the time evolution of the spatial geometry.

In a conventional FRW model, the $3$-dimensional manifold $\Sigma$
is an isotropic and homogeneous Riemannian manifold. More
precisely, $\Sigma$ is globally isotropic. To explain the
significance of this, we shall review the notions of homogeneity
and isotropy.

A Riemannian manifold $(\Sigma,\gamma)$ is defined to be
homogeneous if the isometry group $I(\Sigma)$ acts transitively
upon $\Sigma$. For any pair of points $p,q \in \Sigma$ from a
homogeneous manifold, there will be an isometry $\phi$ such that
$\phi(p)=q$. If there is a unique isometry $\phi$ such that
$\phi(p)=q$ for each pair of points $p,q \in \Sigma$, then the
isometry group action is said to be simply transitive. If there is
sometimes, or always, more than one such isometry, then the
isometry group action is said to be multiply transitive.

In colloquial terms, one can say that the geometrical
characteristics at one point of a homogeneous Riemannian manifold,
match those at any other point.

To define isotropy, it is necessary to introduce the `isotropy
subgroup'. At each point $p \in \Sigma$ of a Riemannian manifold,
there is a subgroup $H_p \subset I(\Sigma)$ of the isometry group.
Referred to as the isotropy subgroup at $p$, $H_p$ is the set of
isometries under which $p$ remains fixed. Thus, $\psi \in H_p$ is
such that $\psi(p)=p$. The differential map $\psi_*$ of each $\psi
\in H_p$, bijectively maps the tangent space at $p$ onto itself.
By restricting the differential map $\psi_*$ of each $\psi \in
H_p$ to $T_p \Sigma$, the tangent space at $p$, one obtains a
linear representation of the isotropy subgroup $H_p$:

$$
j:H_p \rightarrow GL(T_p \Sigma)\,.
$$

We can refer to $j(H_p)$ as the linear isotropy subgroup at $p$.
Whilst $H_p$ is a group of transformations of $\Sigma$, $j(H_p)$
is a group of transformations of $T_p \Sigma$.

The Riemannian metric tensor field $\gamma$ upon the manifold
$\Sigma$, assigns a positive-definite inner product $\langle \; ,
\; \rangle_\gamma$ to each tangent vector space $T_p \Sigma$.
Hence, each tangent vector space can be considered to be an inner
product space

$$
(T_p \Sigma, \langle \; , \; \rangle _\gamma)\,.
$$

Whilst $H_p$ is a group of diffeomorphic isometries of the
Riemannian manifold $(\Sigma,\gamma)$, $j(H_p)$ is a group of
linear isometries of the inner product space $(T_p \Sigma,\langle
\; , \; \rangle_\gamma)$. For any pair of vectors $v,w \in T_p
\Sigma$, and for any $\psi \in j(H_p)$, this means that

$$
\langle \psi(v),\psi(w) \rangle = \langle v,w \rangle \,.
$$

We can therefore consider the representation $j$ to be an
orthogonal linear representation:

$$
j:H_p \rightarrow O(T_p \Sigma) \subset GL(T_p \Sigma) \,.
$$

We can now define a Riemannian manifold $(\Sigma,\gamma)$ to be
isotropic at a point $p$ if the linear isotropy group at $p$,
$j(H_p)$, acts transitively upon the unit sphere in the tangent
space $T_p \Sigma$.

This definition requires some elaboration. Firstly, the unit
sphere $S_p \Sigma \subset T_p \Sigma$ is defined as

$$
S_p \Sigma = \{v \in T_p \Sigma : \langle v,v \rangle_\gamma = 1
\} \,.
$$

The unit sphere represents all possible directions at the point
$p$ of the manifold $\Sigma$. Each vector $v \in S_p \Sigma$ can
be considered to point in a particular direction.

Now, the requirement that $j(H_p)$ acts transitively upon $S_p
\Sigma$, means that for any pair of points $v,w \in S_p \Sigma$ on
the unit sphere, there must be a linear isometry $\psi \in j(H_p)$
such that $\psi(v)=w$. If $j(H_p)$ acts transitively upon the unit
sphere $S_p \Sigma$, all directions at the point $p$ are
geometrically indistinguishable. If a Riemannian manifold
$(\Sigma,\gamma)$ is isotropic at a point $p$, then all directions
at the point $p$ are geometrically indistinguishable.

In the case of cosmological relevance, where $(\Sigma,\gamma)$ is
a $3$-dimensional Riemannian manifold which represents the spatial
universe, isotropy at a point $p$ means that all spatial
directions at $p$ are indistinguishable.

It is simple to show that $j(H_p)$ acts transitively upon the unit
sphere at a point $p$, if and only if it acts transitively upon a
sphere of any radius in $T_p \Sigma$. Hence, if $(\Sigma,\gamma)$
is isotropic at $p$, then $j(H_p)$ includes the so-called rotation
group $SO(T_p \Sigma) \cong SO(3)$. The orbits of the action are
the concentric family of $2$-dimensional spheres in $T_p \Sigma$,
plus the single point at the origin of the vector space.

If $j(H_p)$, the linear isotropy group at $p$, acts transitively
upon the unit sphere in $T_p \Sigma$, then each orbit of the
isotropy group action on $\Sigma$ consists of the points which lie
a fixed distance from $p$, and each such orbit is a homogeneous
surface in $\Sigma$, whose isometry group contains $SO(3)$.

An isotropic Riemannian manifold $(\Sigma,\gamma)$ is defined to
be a Riemannian manifold which is isotropic at every point $p \in
\Sigma$. To be precise, we have defined a globally isotropic
Riemannian manifold. We will subsequently introduce the notion of
local isotropy, which generalises the notion of global isotropy.
It is conventionally understood that when one speaks of isotropy,
one is speaking of global isotropy unless otherwise indicated. To
clarify the discussion, however, we will hereafter speak
explicitly of global isotropy.

From the perspective of the $4$-dimensional Lorentzian manifold
$\mathcal{M} = I \times_R \Sigma$, each point $p$ belongs to a
spacelike hypersurface $\Sigma_t = t \times \Sigma$ which is
isometric with $(\Sigma, R(t)^2\gamma)$. The hypersurface
$\Sigma_t$ is a $3$-dimensional Riemannian manifold of constant
sectional curvature. The tangent space $T_p\mathcal{M}$ contains
many $3$-dimensional spacelike subspaces, but only one,
$T_p\Sigma_t$, which is tangent to $\Sigma_t$, the hypersurface of
constant sectional curvature passing through $p$. The unit sphere
$S_p\Sigma_t$ in this subspace represents all the possible spatial
directions at $p$ in the hypersurface of constant sectional
curvature. Spatial isotropy means that the isotropy group at $p$
acts transitively upon this sphere. Whilst there is a spacelike
unit sphere in each $3$-dimensional spacelike subspace of
$T_p\mathcal{M}$, the spatial isotropy of the FRW models pertains
only to the transitivity of the isotropy group action upon
$S_p\Sigma_t$. However, there is also a null sphere at $p$
consisting of all the null lines in $T_p\mathcal{M}$. This sphere
represents all the possible light rays passing through $p$.
Letting $\partial_t$ denote the unit timelike vector tangent to
the one-dimensional submanifold $I$, the isotropy group action at
each $p$ maps $\partial_t$ to itself. Any vector in
$T_p\mathcal{M}$ can be decomposed as the sum of a multiple of
$\partial_t$ with a vector in the spacelike subspace
$T_p\Sigma_t$. Hence, the action of the isotropy group upon
$T_p\Sigma_t$ can be extended to an action upon the entire tangent
vector space $T_p\mathcal{M}$. In particular, the isotropy group
action can be extended to the null sphere. If the isotropy group
action is transitive upon the set of spatial directions
$S_p\Sigma_t$, then it will also be transitive upon the null
sphere at $p$.

In a conventional FRW model, the complete and connected
$3$-dimensional Riemannian manifold $(\Sigma,\gamma)$ is both
homogeneous and globally isotropic. In fact, any connected
$3$-dimensional globally isotropic Riemannian manifold must be
homogeneous. It is therefore redundant to add that a conventional
FRW model is spatially homogeneous.

Now, a complete, connected, globally isotropic $3$-dimensional
Riemannian manifold must be of constant sectional curvature $k$. A
complete, connected Riemannian manifold of constant sectional
curvature, of any dimension, is said to be a Riemannian space
form.

There exists a simply connected, $3$-dimensional Riemannian space
form for every possible value, $k$, of constant sectional
curvature.

\begin{theorem} A complete, simply connected, $3$-dimensional
Riemannian manifold of constant sectional curvature $k$, is
isometric to
\begin{itemize}

\item The sphere $S^3(r) \; \text{for} \; r=\surd(1/k) \; \text{if} \; k>0$

\item Euclidean space $\mathbb{R}^3  \; \text{if} \; k=0$

\item The hyperbolic space $H^3(r) \; \text{for} \; r= \surd(1/-k) \; \text{if} \; k<0$

\end{itemize}
where

    $$S^3(r) = \{x \in \mathbb{R}^4 : \; \langle x,x \rangle = r^2\}$$ and

    $$H^3(r) = \{x \in \mathbb{R}^{3,1} :\;  \langle x,x \rangle = -r^2, \; x^0 >0\} \,.$$

\end{theorem}

$S^3(r)$, the sphere of radius $r$, is understood to have the
metric tensor induced upon it by the embedding of $S^3(r)$ in the
Euclidean space $\mathbb{R}^4$, and the hyperboloid $H^3(r)$ is
understood to have the metric tensor induced upon it by the
embedding of $H^3(r)$ in the pseudo-Euclidean space
$\mathbb{R}^{3,1}$.

Geometries which differ from each other by a scale factor are said
to be homothetic. Space forms are homothetic if and only if their
sectional curvature is of the same sign. There are, therefore, up
to homothety, only three $3$-dimensional simply connected
Riemannian space forms: $S^3$, the three-dimensional sphere;
$\mathbb{R}^3$, the three-dimensional Euclidean space; and $H^3$,
the three-dimensional hyperboloid.

Whilst it is true that every simply connected Riemannian space
form is globally isotropic, the converse is not true.
Real-projective three-space $\mathbb{R}\mathbb{P}^3$, equipped
with its canonical metric tensor, is also globally isotropic, but
is non-simply connected.

Up to homothety, there are four possible spatial geometries of a
conventional, globally isotropic FRW model: $S^3$, $\mathbb{R}^3$,
$H^3$, and $\mathbb{RP}^3$. Up to homothety, these are the only
complete and connected, globally isotropic $3$-dimensional
Riemannian manifolds, (Beem and Ehrlich 1981, p131).

$\mathbb{R}^3$ and $H^3$ are diffeomorphic, hence there are only
three possible spatial topologies of a globally isotropic FRW
model. Only $S^3$, $\mathbb{R}^3$, and $\mathbb{RP}^3$ can be
equipped with a globally isotropic, complete Riemannian metric
tensor.

A generalisation of the conventional FRW models can be obtained by
dropping the requirement of global isotropy, and substituting in
its place the condition that $(\Sigma,\gamma)$ must be a
Riemannian manifold of constant sectional curvature, a space form.

As already stated, every globally isotropic Riemannian
$3$-manifold is a space form. However, not every $3$-dimensional
Riemannian space form is globally isotropic. On the contrary,
there are \emph{many} $3$-dimensional Riemannian space forms which
are not globally isotropic.

One can obtain any $3$-dimensional Riemannian space form as a
quotient $\Sigma/\Gamma$ of a simply connected Riemannian space
form, where $\Gamma$ is a discrete, properly discontinuous,
fixed-point free subgroup of the isometry group $I(\Sigma)$,
(O'Neill 1983, p243 and Boothby 1986, p406, Theorem 6.5). Properly
discontinuous means that for any compact subset $C \subset
\Sigma$, the set $\{\phi \in \Gamma: \; \phi(C) \cap C \neq
\emptyset \}$ is finite. The quotient is guaranteed to be
Hausdorff if the action is properly discontinuous. $\Gamma$ acts
properly discontinuously if and only if $\Gamma$ is a discrete
group, hence there is some redundancy in the definition above.

The quotient $\Sigma/\Gamma$ is a Riemannian manifold if and only
if $\Gamma$ acts freely. The natural way of rendering the quotient
manifold $\Sigma/\Gamma$ a Riemannian manifold ensures that
$\Sigma$ is a Riemannian covering of $\Sigma/\Gamma$, (see O'Neill
1983, p191, for a general version of this where $\Sigma$ is a
semi-Riemannian manifold). The covering map $\eta:\Sigma
\rightarrow \Sigma/\Gamma$ is a local isometry, hence if $\Sigma$
is of constant sectional curvature $k$, then $\Sigma/\Gamma$ will
also be of constant sectional curvature $k$. If $\Sigma$ is simply
connected, then the fundamental group of the quotient manifold
$\Sigma/\Gamma$ will be isomorphic to $\Gamma$. i.e.
$\pi_1(\Sigma/\Gamma) = \Gamma$. Hence, for a non-trivial group
$\Gamma$, the quotient manifold will not be simply connected.

Every space form of constant sectional curvature $k > 0$ is a
quotient $S^3(r)/\Gamma$, every $k=0$ space form is a quotient
$\mathbb{R}^3/\Gamma$, and every $k < 0$ space form is a quotient
$H^3(r)/\Gamma$.

Let $N(\Gamma)$ denote the normalizer of $\Gamma$ in $I(\Sigma)$.
$N(\Gamma)$ is the largest subgroup of $I(\Sigma)$ which contains
$\Gamma$ as a normal subgroup. The isometry group of
$\Sigma/\Gamma$ is $N(\Gamma)/\Gamma$, (O'Neill 1983, p249).
Equivalently, the isometry group of the quotient manifold is the
`centralizer', or `commutant' $Z(\Gamma)$, the subgroup of
$I(\Sigma)$ consisting of elements which commute with all the
elements of $\Gamma$, (Ellis 1971, p11). In general, there is no
reason for $Z(\Gamma)$ to contain $\Gamma$ as a subgroup.

If $\Gamma$ is a discrete group acting freely on a manifold
$\mathcal{M}$, then there is a `fundamental cell' $C \subset
\mathcal{M}$, a closed subset, whose images under $\Gamma$
tessellate the space $\mathcal{M}$. Each orbit $\Gamma x$, for $x
\in \mathcal{M}$, contains either one interior point of $C$, or
two or more boundary points of $C$. A fundamental cell therefore
contains representatives of each orbit of $\Gamma$, and for almost
all orbits, the fundamental cell contains exactly one
representative. Given that a point of $\mathcal{M}/\Gamma$ is an
orbit of $\Gamma$, it follows that one can construct
$\mathcal{M}/\Gamma$ from $C$ by identifying boundary points in
the same orbit, (J.L.Friedman 1991, p543-545).

To say that the subsets $\{\phi(C): \phi \in \Gamma \}$,
tessellate the space $\mathcal{M}$, means that they provide a
covering of $\mathcal{M}$ by isometric closed subsets, no two of
which have common interior points.

In the case of a quotient $\Sigma/\Gamma$ of a simply connected
$3$-dimensional Riemannian space form $\Sigma$, by a discrete
group $\Gamma$ of properly discontinuous, freely acting
isometries, if $\Sigma/\Gamma$ is a compact $3$-manifold, then the
fundamental cell is a polyhedron, (J.L.Friedman 1991, p544). One
can construct such compact quotient manifolds by identifying the
faces of the polyhedron. The best-known example is the way in
which one can obtain the three-torus $\mathbb{T}^3$ by identifying
the opposite faces of a cube.

Although many space forms are not globally isotropic, they are
all, at the very least, locally isotropic. To define local
isotropy, it is necessary to use the concept of a local isometry.
One can define a local isometry of a Riemannian manifold
$(\Sigma,\gamma)$ to be a smooth map $\phi:\Sigma \rightarrow
\Sigma$, such that each differential map $\phi_{*p}:T_p \Sigma
\rightarrow T_{\phi(p)}\Sigma$ is a linear isometry. Equivalently,
the defining characteristic of a local isometry is that each $p
\in \Sigma$ has a neighbourhood $V$ which is mapped by $\phi$ onto
an isometric neighbourhood $\phi(V)$ of $\phi(p)$. Whilst a local
isometry $\phi:\Sigma \rightarrow \Sigma$ need not be a
diffeomorphism of $\Sigma$, it must at the very least be a local
diffeomorphism. It is also worth noting that every isometry must
be a local isometry.

At each point $p \in \Sigma$ of a Riemannian manifold, one can
consider the family of all local isometries of $(\Sigma,\gamma)$
which leave the point $p$ fixed. Each such local isometry maps a
neighbourhood $V$ of $p$ onto an isometric neighbourhood of the
same point $p$. This family of local isometries is the analogue of
the isotropy subgroup at $p$ of the global isometry group. The
differential map of each such local isometry, $\phi_{*p}: T_p
\Sigma \rightarrow T_p \Sigma$, is a linear isometry of the inner
product space $(T_p \Sigma,\langle \; , \; \rangle_\gamma)$.

One defines a Riemannian manifold $(\Sigma,\gamma)$ to be locally
isotropic at a point $p$ if the family of local isometries which
leave $p$ fixed, act transitively upon the unit sphere $S_p \Sigma
\subset T_p \Sigma$. If the local linear isotropy group at $p$
acts transitively upon $S_p \Sigma$, then the local isotropy group
at $p$ must contain $SO(3)$. Naturally, a locally isotropic
Riemannian manifold is defined to be a Riemannian manifold which
is locally isotropic at every point.

Outside the common neighbourhood $U$ of the local isometries in
the local isotropy group at $p$, the set of points which lie at a
fixed spatial distance from $p$, will not, in general, form a
homogeneous surface. If a locally isotropic space has been
obtained as the quotient of a discrete group action, then beyond
the neighbourhood $U$, the set of points which lie at a fixed
distance from $p$, will, in general, have a discrete isometry
group. Inside $U$, the set of points which lie at a fixed spatial
distance from $p$, will still form a homogeneous surface with an
isometry group that contains $SO(3)$.

Beyond the neighbourhood $U$, the orbits of the local isotropy
group of $p$ still coincide with sets of points that lie at a
fixed distance from $p$, but, to reiterate, these orbits are not
homogeneous surfaces. The local isotropy group of $p$, which
contains $SO(3)$, acts transitively upon these surfaces, but it
does not act as a group of isometries upon these surfaces.
Instead, only a discrete subgroup of $SO(3)$ provides the
isometries of these surfaces.

Beyond $U$, the orbits of the local isotropy group action have
preferred directions. If a locally isotropic space has been
obtained as the quotient of a discrete isometry group action, and
if that quotient action is defined by identifying the faces of a
polyhedron, then, beyond a certain distance from each point $p$,
the perpendiculars to the faces of the polyhedron define preferred
directions on the orbits of the local isotropy group action.

Projecting from the hypersurfaces $\Sigma_t$ of a warped product
$I \times_R \Sigma$ onto the $3$-dimensional locally isotropic
Riemannian manifold $\Sigma$, the past light cone $E^-(x)$ of an
arbitrary point $x = (t_0,p)$, passes through the orbits in
$\Sigma$ of the local isotropy group of $p \in \Sigma$ at
ever-greater distances from $p$ the further the light cone reaches
into the past\footnote{The past light cone $E^-(x)$ is the set of
points which can be connected to $x$ by a future directed null
curve.}. Hence, in a locally isotropic warped product, the past
light cone $E^-(x)$ will consist of homogeneous $2$-dimensional
surfaces close to $x$, but beyond a certain spatial distance,
looking beyond a certain time in the past, the constant time
sections of the light cone will consist of non-homogeneous
surfaces, which only have a discrete isometry group.
Observationally, this means that one would only see an isotropic
pattern of light sources up to a certain distance, or up to a
certain `look-back' time, away from the point of observation.

It is easy to see that every globally isotropic Riemannian
manifold must be locally isotropic. However, there are many
locally isotropic Riemannian manifolds which are not globally
isotropic. Whilst every Riemannian space form is locally
isotropic, only a simply connected space form is guaranteed to be
globally isotropic.

Present astronomical data indicates that the spatial universe is
locally isotropic about our location in space. Present data does
not reveal whether the spatial universe is globally isotropic
about our point in space. We have only received light from a
proper subset of the spatial universe because light from more
distant regions has not had time to reach us.

The Copernican Principle declares that the perspective which the
human race has upon the universe is highly typical. Combining this
philosophical principle with the astronomical evidence that the
spatial universe is locally isotropic about our point in space,
one infers that the spatial universe is locally isotropic about
every point in space. One infers that the spatial universe is
representable by a locally isotropic Riemannian manifold. Our
limited astronomical data means that it is unjustified to
stipulate that the spatial universe is globally isotropic.

Neither do our astronomical observations entail global
homogeneity; we only observe local homogeneity, and approximate
local homogeneity at that. One can define a Riemannian manifold
$(\Sigma,\gamma)$ to be locally homogeneous if and only if, for
any pair of points $p,q \in \Sigma$, there is a neighbourhood $V$
of $p$, which is isometric with a neighbourhood $U$ of $q$. There
will be a local isometry $\phi:\Sigma \rightarrow \Sigma$ such
that $\phi(V)=U$. Just as all of the space forms are locally
isotropic, so they are also locally homogeneous. A connected,
globally isotropic Riemannian manifold must be globally
homogeneous, and similarly, a connected locally isotropic
Riemannian manifold must be locally homogeneous, (Wolf 1967,
p381-382).

This generalisation of the conventional FRW models enlarges the
range of possible spatial geometries and topologies of our
universe. The topology of the spatial universe need not be
homeomorphic to either $\mathbb{R}^3$, $S^3$, or $\mathbb{RP}^3$.

Take the $3$-dimensional Euclidean space forms. These are the
complete, connected, flat $3$-dimensional Riemannian manifolds,
each of which is a quotient $\mathbb{R}^3/\Gamma$ of
$3$-dimensional Euclidean space by a discrete group $\Gamma$ of
properly discontinuous, fixed point free isometries.

There are actually $18$ non-homeomorphic $3$-dimensional manifolds
which can be equipped with a complete Riemannian metric tensor of
constant sectional curvature $k=0$. Of the $18$ there are $10$ of
compact topology, and $8$ of non-compact topology. The non-compact
$3$-dimensional Euclidean space forms include, (Wolf 1967,
p112-113):

\begin{enumerate}
\item $\mathbb{R}^1 \times \mathbb{R}^2 \equiv \mathbb{R}^3$
\item $\mathbb{R}^1 \times \text{(Cylinder)}$
\item $\mathbb{R}^1 \times \text{(Torus) i.e. } \mathbb{R}^1 \times \mathbb{T}^2$
\item $\mathbb{R}^1 \times \text{(Moebius band)}$
\item $\mathbb{R}^1 \times \text{(Klein bottle)}$
\end{enumerate}

The second factors in the five cases listed above, exhaust the
$2$-dimensional Euclidean space forms. Because the Moebius band
and the Klein bottle are non-orientable, cases 4. and 5. are
non-orientable $3$-manifolds. The first three cases are, however,
orientable.

Of the $8$ non-compact Euclidean space forms, $4$ are orientable,
and $4$ are non-orientable. Of the $10$ compact Euclidean space
forms, $6$ are orientable, and $4$ are non-orientable.

Notice that $\mathbb{R}^1 \times \text{(Cylinder)} \equiv
\mathbb{R}^1 \times (\mathbb{R}^1 \times S^1) \cong \mathbb{R}^2
\times S^1$. The open disc $B^2$ is homeomorphic with
$\mathbb{R}^2$, hence $\mathbb{R}^2 \times S^1 \cong B^2 \times
S^1$. The $3$-manifold $B^2 \times S^1$ is the interior of a solid
torus. The interior of a solid torus is a possible spatial
topology for a FRW universe. All our astronomical observations, in
conjunction with the Copernican principle, are consistent with the
spatial universe having the shape of a solid ring.

The compact $3$-dimensional Euclidean space forms include the
$3$-dimensional torus $\mathbb{T}^3$. The other nine compact flat
Riemannian manifolds can then be obtained from $\mathbb{T}^3$ as a
quotient $\mathbb{T}^3/\Gamma$, (Wolf 1967, p105). Of the ten
compact flat Riemannian $3$-manifolds, nine can be fibred over the
circle. In seven of these cases, the fibre is a $2$-torus, and in
the other two cases, the fibre is a Klein bottle, (Besse 1987,
p158, 6.20). Alternatively, one can treat two of the ten as circle
bundles over the Klein bottle, and one of these is the trivial
product bundle $S^1 \times (\text{Klein bottle})$, (Besse 1987,
p158, 6.19).

\hfill \break

In the case of the $3$-dimensional Riemannian space forms of
positive curvature $S^3/\Gamma$, the isometry group of $S^3$ is
$SO(4)$, and the quotient group $\Gamma$ must be a discrete
subgroup of $SO(4)$ which acts freely and discontinuously on
$S^3$. These subgroups come in three types (Rey and Luminet 2003,
p52-55; Wolf 1967, p83-87): $\mathbb{Z}_p$ the cyclic rotation
groups of order $p$, for $p \geq 2$; $D_m$ the dihedral groups of
order $2m$, for $m > 2$, the symmetry groups of the regular
$m$-sided polygons; and the symmetry groups of the regular
polyhedra, $T$, $O$ and $I$. There are actually five regular
polyhedra (the `Platonic solids'): the regular tetrahedron (4
faces), the regular hexahedron or cube (6 faces), the regular
octahedron (8 faces), the regular dodecahedron (12 faces), and the
regular icosahedron (20 faces). There are, however, only three
distinct symmetry groups, the tetrahedral group $T$, octahedral
group $O$, and icosahedral group $I$. The hexahedron has the
octahedral symmetry group $O$, and the dodecahedron has the
icosahedral symmetry group $I$. There are also double coverings of
the dihedral and polyhedral groups, denoted as
$D_m^*,T^*,O^*,I^*$.

The \emph{globally} homogeneous $3$-dimensional Riemannian space
forms of positive curvature $S^3/\Gamma$ can be listed as follows,
(Wolf 1967, p89, Corollary 2.7.2):

\begin{enumerate}
\item $S^3$
\item $\mathbb{RP}^3 \cong S^3/\mathbb{Z}_2$
\item $S^3/\mathbb{Z}_p  \; \text{for} \; p > 2$
\item $S^3/D_m^* \; \text{for} \; m > 2$
\item $S^3/T^*$
\item $S^3/O^*$
\item $S^3/I^*$
\end{enumerate}

One can also take the quotient of $S^3$ with respect to groups
$\Gamma$ of the form $\mathbb{Z}_u \times D^*_v$, $\mathbb{Z}_u
\times T^*_v$, $\mathbb{Z}_u \times O^*$, or $\mathbb{Z}_u \times
I^*$, for certain values of $u$ and $v$, and where the $T^*_v$ are
subgroups of $T^*$ (Ellis 1971 p13). These spherical space forms
are merely locally homogeneous.

Note that, in contrast with the Euclidean case, there are an
infinite number of distinct spherical space forms because there is
no limit on $p$ or $m$.

The real projective space $\mathbb{RP}^3 \cong S^3/\mathbb{Z}_2$
is globally isotropic and orientable, but not simply connected.
The spaces $S^3/\mathbb{Z}_p$ are referred to as lens spaces,
while the Poincare manifold is homeomorphic with $S^3/I$.

All of the $3$-dimensional spherical Riemannian space forms are of
compact topology.

\hfill \break

In the case of the $3$-dimensional hyperbolic space forms, the
work of Thurston demonstrates that `most' compact and orientable
$3$-manifolds can be equipped with a complete Riemannian metric
tensor of constant negative sectional curvature. This means that
`most' compact, orientable $3$-manifolds can be obtained as a
quotient $H^3/\Gamma$ of hyperbolic $3$-space. The meaning of
`most' in this context involves Dehn surgery, (Besse 1987,
p159-160).

Every compact, orientable $3$-manifold can be obtained from $S^3$
by Dehn surgery along some link $L$. A link in a manifold is
defined to be a finite, disjoint union of simple closed curves $L
= J_1 \cup \cdots \cup J_n$. The first step of Dehn surgery along
a given link $L$, is to specify disjoint tubular neighbourhoods
$N_i$ of each component $J_i$. Each tubular neighbourhood $N_i$ is
homeomorphic with a solid torus $D^2 \times S^1$.

Having identified $n$ disjoint solid tori in $S^3$, some of which
may be knotted, one removes the interior $\text{Int}(N_i)$ of
each. That is, one takes the complement

$$
S^3 - (\text{Int}(N_1) \cup \cdots \cup \text{Int}(N_n)) \,.
$$The boundary surface $\partial N_i$ of the hole left by the
removal of $\text{Int}(N_i)$ is homeomorphic with a
$2$-dimensional torus $\mathbb{T}^2$. Thus, what remains is a
manifold bounded by $n$ disjoint $2$-dimensional tori.

Next, one takes $n$ copies of the solid torus $M_i$, and one sews
each solid torus back into $ S^3 - (\text{Int}(N_1) \cup \cdots
\cup \text{Int}(N_n))$. Each sewing instruction is specified by a
diffeomorphism

$$
\phi_i:\partial M_i \rightarrow \partial N_i \,.
$$ One defines a point $x \in \partial M_i$ on the boundary of the
solid torus $\partial M_i$ to be equivalent to the point
$\phi_i(x) \in \partial N_i$ on the boundary of the hole left by
the removal of $\text{Int}(N_i)$. The result is a new
$3$-manifold.

By varying the choice of link, and by varying the choice of sewing
instructions, one can obtain every compact, orientable
$3$-manifold. Furthermore, one can obtain every compact,
orientable $3$-manifold even if one limits the Dehn surgery to
hyperbolic links. A link $L$ in $S^3$ is defined to be a
hyperbolic link if $S^3 - L$ can be equipped with a complete
Riemannian metric tensor of constant negative sectional curvature.

Given a choice of link $L$, although it is true that some
collections of diffeomorphisms $\{\phi_i:\partial M_i \rightarrow
\partial N_i : \, i=1,...,n \}$ yield the same manifold, there are
still an uncountable infinity of distinct ways in which one can
sew the solid tori back in. Thurston has shown that, in the case
of a hyperbolic link, only a finite number of choices for the
sewing instructions yield a manifold which cannot support a
complete Riemannian metric tensor of constant negative sectional
curvature. It is in this sense that `most' compact, orientable
$3$-manifolds can be equipped with a complete Riemannian metric
tensor of constant negative sectional curvature. Given that all
the hyperbolic $3$-dimensional Riemannian space forms can be
obtained as quotients $H^3/\Gamma$, it follows that `most'
compact, orientable $3$-manifolds can be obtained as such a
quotient.

The cosmological corollary of Thurston's work is that there exists
a vast class of compact, orientable $3$-manifolds, which could
provide the topology of a $k < 0$ FRW universe. Note that there
are compact and non-compact quotients $H^3/\Gamma$.

The rigidity theorem for hyperbolic space-forms states that a
connected oriented $n$-dimensional manifold, compact or
non-compact, of dimension $n \geq 3$, supports at most one
Riemannian metric tensor of constant negative sectional curvature,
\textit{up to homothety}. Unfortunately, this last qualification
has been neglected in some places, and misunderstanding has
resulted amongst cosmologists. Given any $3$-dimensional
Riemannian space form $\Sigma$ of constant negative curvature $k$,
one can change the geometry by an arbitrary scale factor $f$ to
obtain a Riemannian manifold with the same topology, but with
constant negative curvature $k/f^2$. This, after all, is what the
time-dependent scale factor does with a hyperbolic universe in FRW
cosmology!\footnote{A scale factor $f$ used to define the
Riemannian geometry $(\Sigma,\gamma)$ should not be confused with
the dynamic scale factor $R(t)$ of a FRW universe.} Cornish and
Weeks falsely state that if a pair of $3$-dimensional hyperbolic
manifolds are homeomorphic, then they must be isometric, (Cornish
and Weeks 1998, p8). Rey and Luminet, (2003, p57-58), state that,
for $n \geq 3$, a connected oriented $n$-manifold can support at
most one hyperbolic metric, without adding the qualification, `up
to homothety'. They falsely state that if two hyperbolic manifolds
of dimension $n \geq 3$ have isomorphic fundamental groups, then
they must be isometric. There is no reason why the metric on a
manifold obtained as a quotient $H^3/\Gamma$ cannot be changed by
a scale factor from the metric it inherits in the quotient
construction. Alternatively, there is no reason why the canonical
metric on the universal cover $H^3$ cannot be changed by a scale
factor before the quotient is taken. This is equivalent to
introducing, as possible universal covers, the homothetic family
of hyperbolic 3-manifolds of radius $r$, $\{H^3(r): r \in
(0,\infty) \}$. The family of quotients $\{H^3(r)/\Gamma: r \in
(0,\infty) \}$ are also mutually homothetic. Each such quotient
$H^3(r)/\Gamma$ possesses the same fundamental group, namely
$\Gamma$, but the family of geometries $\{H^3(r)/\Gamma: r \in
(0,\infty) \}$ are merely homothetic, not isometric. Luminet and
Roukema (1999, p14) correctly state the rigidity theorem with the
vital qualification of a fixed scale factor on the universal
cover.

The rigidity theorem means that for a fixed $\Gamma$, (up to group
isomorphy), and for a fixed scale factor (`radius of curvature')
$r$ on the universal cover $H^3(r)$, there is a unique
$H^3(r)/\Gamma$ up to isometry. In the case of \emph{compact}
hyperbolic 3-manifolds, for fixed $\Gamma$ and $H^3(r)$, the
volume of the fundamental cell in $H^3(r)$, and therefore the
volume of the quotient $H^3(r)/\Gamma$, is unique. Hence, volumes
can be used to classify compact hyperbolic 3-manifolds as long as
one adds the vital qualification that the volumes are expressed in
units of the `curvature radius', i.e. in units of the scale
factor. If one fixes $\Gamma$, but permits $r$ to vary, then the
volume of the quotient can vary arbitrarily. It is the volume
expressed in curvature radius units which is unique, not the
absolute value of the volume. It has also been shown that the
volume of any compact hyperbolic 3-manifold, in curvature radius
units, is bounded from below by $V_{min} = 0.166 r^3$. This is
more a constraint on the relationship between volume and the scale
factor on the universal cover in the quotient construction, than
an absolute lower limit on volume.

These volume constraints, however, are significant because they
are absent in the case of the compact Euclidean 3-manifolds. Even
with the scale factor of the flat metric on $\mathbb{R}^3$ fixed,
and with $\Gamma$ fixed, one can choose fundamental cells of
arbitrary volume in $\mathbb{R}^3$, hence the volume of
$\mathbb{R}^3/\Gamma$ can be varied arbitrarily even when
expressed in units of curvature radius.

\hfill \break

To reiterate, a connected, locally isotropic Riemannian manifold
is only guaranteed to be locally homogeneous. The only
\emph{globally} homogeneous $3$-dimensional Riemannian space forms
are, (Wolf 1967, p88-89):

\begin{enumerate}
\item $k = 0$ \begin{enumerate}
\item $\mathbb{T}^3$
\item $\mathbb{R}^3$
\item $\mathbb{R}^2 \times S^1$
\item $\mathbb{R} \times \mathbb{T}^2$
\end{enumerate}
\item $k > 0$ \begin{enumerate}
\item $S^3$
\item $\mathbb{RP}^3 \cong S^3/\mathbb{Z}_2$
\item $S^3/\mathbb{Z}_p  \; \text{for} \; p > 2$
\item $S^3/D_m^* \; \text{for} \; m > 2$
\item $S^3/T^*$
\item $S^3/O^*$
\item $S^3/I^*$
\end{enumerate}
\item $k < 0$ \begin{enumerate}
\item $H^3$ \end{enumerate}
\end{enumerate}

Clearly, there is no compact, globally homogeneous,
$3$-dimensional hyperbolic space-form because the only globally
homogeneous $3$-dimensional hyperbolic space-form is the
non-compact space $H^3$ (Wolf 1967, p90, Lemma 2.7.4 and p230,
Theorem 7.6.7). Also note that only four of the eighteen
$3$-dimensional Euclidean space-forms are globally homogeneous.

\subsection{The Friedmann equations}

Given a particular Lorentzian metric tensor field $g$, the
Einstein field equation determines the corresponding stress-energy
tensor field $T$. In coordinate-independent notation, the Einstein
field equation, without cosmological constant, can be expressed as

$$
T = 1/(8\pi G)(\text{Ric} - 1/2 \; \text{S} \, g) \,.
$$ $Ric$ denotes the Ricci tensor field determined by $g$,
and $\text{S}$ denotes the curvature scalar field. We have chosen
units here in which $c=1$. In the component terms used by
physicists,

$$
T_{\mu \nu} = 1/(8\pi G)(R_{\mu\nu} - 1/2 \; \text{S} \,
g_{\mu\nu}) \,.
$$

In the Friedmann-Robertson-Walker models, the warped product
metric, $$g = -dt \otimes dt + R(t)^2 \gamma \, ,$$ corresponds to
the stress-energy tensor of a perfect fluid:

$$
T = (\rho + p)dt \otimes dt + pg \,.
$$ $\rho$ and $p$ are both scalar fields on $\mathcal{M}$, constant on each
hypersurface $\Sigma_t = t \times \Sigma$, but time dependent.
$\rho$ is the energy density function, and $p$ is the pressure
function.

The scale factor $R(t)$, energy density $\rho(t)$, and pressure
$p(t)$ of a Friedmann-Robertson-Walker model satisfy the so-called
Friedmann equations, (O'Neill 1983, p346; Kolb and Turner 1990,
p49-50):

$$
\frac{8\pi G}{3}\rho(t) = \left( \frac{R'(t)}{R(t)} \right)^2 +
\frac{k}{R(t)^2} \, ,
$$

$$
-8\pi G p(t) = 2 \frac{R''(t)}{R(t)} + \left( \frac{R'(t)}{R(t)}
\right) ^2 + \frac{k}{R(t)^2} \,.
$$ $k$ is the constant sectional curvature of the $3$-dimensional
Riemannian space form $(\Sigma,\gamma)$; $R'(t) \equiv
\frac{dR(t)}{dt}$; and $R'(t)/R(t)$ is the Hubble parameter
$H(t)$. The sectional curvature of the hypersurface $\Sigma_t$ is
$k/R(t)^2$.

\subsection{The Hubble parameter, redshift, and horizons}

A Riemannian manifold $(\Sigma,\gamma)$ is equipped with a natural
metric space structure $(\Sigma,d)$. In other words, there exists
a non-negative real-valued function $d:\Sigma \times \Sigma
\rightarrow \mathbb{R}$ which is such that

\begin{enumerate}
\item $d(p,q) = d(q,p)$
\item $d(p,q) + d(q,r) \geq d(p,r)$
\item $d(p,q) = 0$ iff $p =q$
\end{enumerate}

The metric tensor $\gamma$ determines the Riemannian distance
$d(p,q)$ between any pair of points $p,q \in \Sigma$. The metric
tensor $\gamma$ defines the length of all curves in the manifold,
and the Riemannian distance is defined as the infimum of the
length of all the piecewise smooth curves between $p$ and $q$. In
the warped product space-time $I \times_R \Sigma$, the spatial
distance between $(t,p)$ and $(t,q)$ is $R(t)d(p,q)$. Hence, if
one projects onto $\Sigma$, one has a time-dependent distance
function on the points of space,

$$
d_t(p,q) = R(t)d(p,q) \,.
$$

Each hypersurface $\Sigma_t$ is a Riemannian manifold
$(\Sigma_t,R(t)^2\gamma)$, and $R(t)d(p,q)$ is the distance
between $(t,p)$ and $(t,q)$ due to the metric space structure
$(\Sigma_t,d_t)$.

The rate of change of the distance between a pair of points in
space is given by

$$\eqalign{
d/dt (d_t(p,q)) &= d/dt (R(t)d(p,q)) \cr &= R'(t)d(p,q) \cr &=
\frac{R'(t)}{R(t)}R(t)d(p,q) \cr &= H(t)R(t)d(p,q) \cr &=
H(t)d_t(p,q)\,. }
$$ The rate of change of distance between a pair of points is
proportional to the spatial separation of those points, and the
constant of proportionality is the Hubble parameter $H(t) =
R'(t)/R(t)$. Galaxies are embedded in space, and the distance
between galaxies increases as a result of the expansion of space,
not as a result of the galaxies moving through space. The rate of
change of the distance between ourselves and a galaxy is referred
to as the recessional velocity $v$ of the galaxy. Where $H_0$
denotes the current value of the Hubble parameter, the Hubble law
is simply $v = H_0d_0$. The recessional velocity corresponds to
the redshift in the spectrum of light received from the galaxy. If
$\lambda_o$ denotes the observed wavelength of light and
$\lambda_e$ denotes the emitted wavelength, the redshift $z$ is
defined as

$$
z = \frac{\lambda_o - \lambda_e}{\lambda_e} =
\frac{\lambda_o}{\lambda_e} - 1 \,.
$$ The distance between ourselves and a galaxy is inferred from a
knowledge of the absolute luminosity of `standard candles' in the
galaxy, and the observed apparent luminosity of those standard
candles.

Cosmology texts often introduce what they call `comoving' spatial
coordinates $(\theta,\phi,r)$. In these coordinates, galaxies
which are not subject to proper motion due to local
inhomogeneities in the distribution of matter, retain the same
spatial coordinates at all times. In effect, comoving spatial
coordinates are merely coordinates upon $\Sigma$ which are lifted
to $I \times \Sigma$ to provide spatial coordinates upon each
hypersurface $\Sigma_t$. The radial coordinate $r$ of a point $q
\in \Sigma$ is chosen to coincide with the Riemannian distance in
the metric space $(\Sigma,d)$ which separates the point at $r=0$
from the point $q$. Hence, assuming the point $p$ lies at the
origin of the comoving coordinate system, the distance between
$(t,p)$ and $(t,q)$ can be expressed in terms of the comoving
coordinate $r(q)$ as $R(t)r(q)$.

If light is emitted from a point $(t_e,p)$ of a warped product
space-time and received at a point $(t_0,q)$, then the integral,

$$
\int^{t_0}_{t_e}\frac{c}{R(t)} \, dt \, ,
$$ where $c$ is the speed of light, expresses the Riemannian
distance $d(p,q)$ in $\Sigma$ travelled by the light between the
point of emission and the point of reception. The present spatial
distance between the point of emission and the point of reception
is

$$
R(t_0)d(p,q) = R(t_0) \int^{t_0}_{t_e}\frac{c}{R(t)} \, dt \,.
$$

The distance which separated the point of emission from the point
of reception at the time the light was emitted is

$$
R(t_e)d(p,q) = R(t_e) \int^{t_0}_{t_e}\frac{c}{R(t)} \, dt \,.
$$
The following integral defines the maximum distance in
$(\Sigma,\gamma)$ from which one can receive light by the present
time $t_0$:
$$
d_{max}(t_0) = \int^{t_0}_{0}\frac{c}{R(t)} \, dt \,.
$$ From this, cosmologists define something called the `particle horizon',

$$
R(t_0) d_{max}(t_0) = R(t_0) \int^{t_0}_{0}\frac{c}{R(t)} \, dt
\,.
$$ We can only receive light from sources which are presently
separated from us by, at most, $R(t_0) d_{max}(t_0)$. In other
words, we can see the past states of luminous objects which are
presently separated from us by a distance of up to $R(t_0)
d_{max}(t_0)$, but we cannot see any further into the spatial
volume of the universe. $R(t_0) \int^{t_0}_{0} c/R(t) \, dt$ is
the present radius of the observable spatial universe.

\subsection{The critical density}

The critical density $\rho_c(t)$ in a FRW model is defined to be
$\rho_c(t) \equiv 3H(t)^2/8\pi G$, and the ratio of the density to
the critical density $\Omega(t) \equiv \rho(t)/\rho_c(t)$ is of
great observational significance. It follows from the Friedmann
equation for the density $\rho$ that one can infer the sign of the
spatial curvature $k$ from $\Omega$. Divide each side of the
Friedmann equation,

$$
\frac{8\pi G}{3}\rho(t) = \left( \frac{R'(t)}{R(t)} \right)^2 +
\frac{k}{R(t)^2} =  H(t)^2 + \frac{k}{R(t)^2} \, ,
$$ by $H(t)^2$ to obtain

$$
\frac{8\pi G}{3}\frac{\rho(t)}{H(t)^2} = 1 + \frac{k}{H(t)^2
R(t)^2} \,.
$$ Now, given that $\Omega(t) = \rho(t)/\rho_c(t)$ it follows that

$$
\Omega(t) = \frac{8\pi G}{3}\frac{\rho(t)}{H(t)^2}\, ,
$$ and one obtains

$$
\Omega(t) - 1 = \frac{k}{H(t)^2 R(t)^2} \,.
$$ Assuming $H(t)^2 R(t)^2 \geq 0$ at the present time, the sign of $k$
must match the sign of $\Omega(t) - 1$ at the present time, (Kolb
and Turner 1990, p50).

If one can infer the current value of the Hubble parameter $H_0$
from observations, one can calculate the current value of the
critical density $\rho_c = 3H_0^2/8\pi G$. If one can also infer
the current average density of matter and energy $\rho_0$ from
observations, then one can calculate $\Omega_0 = \rho_0/\rho_c$.
If $\Omega_0 > 1$, then $k > 0$, if $\Omega_0 = 1$, then $k = 0$,
and if $\Omega_0 < 1$, then $k < 0$.

A FRW universe in which the observed density of matter and energy
is found to be greater than the critical density, must have
spatial curvature $k > 0$, and must be of compact spatial
topology. A FRW universe in which the observed density of matter
and energy is found to equal the critical density, could have the
topology of any one of the $18$ flat $3$-dimensional Riemannian
manifolds. A FRW universe in which the observed density of matter
and energy is found to be less than the critical density, must
have spatial curvature $k < 0$, and could have the topology of any
one of the vast family of 3-manifolds which can be equipped with a
metric tensor of constant negative sectional curvature.

Given a FRW universe, if $\Omega_0 > 1$ then the universe will
exist for a finite time, reaching a maximum diameter before
contracting to a future singularity; if $\Omega_0 = 1$, then the
universe will expand forever, but the expansion rate will converge
to zero, $R'(t) \rightarrow 0 \; \text{as} \; t \rightarrow
\infty$; and if $\Omega_0 < 1$ then the universe will expand
forever.

\subsection{Small universes}

It is commonly assumed in observational cosmology that the
observable spatial universe has the topology of a solid ball
$B^3$, and approximately Euclidean geometry. This assumption could
be derived from the further assumption that the spatial universe
is $\mathbb{R}^3$, with curvature $k=0$, but this would amount to
the selection of a very special geometry. In this context, Blau
and Guth point out, (1987, p532), that $k=0$ is a subset of
measure zero on the real line. As it stands, this is a slightly
glib comment. $k=0$ corresponds to all the flat space forms, not
just $\mathbb{R}^3$, and one requires a justification for placing
a measure on the set of space forms which is derived from a
measure on the set of their sectional curvature values.

It is widely believed that the solid ball topology and approximate
Euclidean geometry of our local spatial universe can be derived
from the assumption that the entire spatial universe is very much
larger than the observable spatial universe. However, this
assumption is not necessarily true, and, moreover, even if it is
true, it does not entail solid ball topology and approximate
Euclidean geometry for our local universe.

Suppose that the spatial universe is compact. A compact Riemannian
manifold $(\Sigma,\gamma)$ is a metric space of finite diameter.
(The diameter of a metric space is the supremum of the distances
which can separate pairs of points). If our universe is a FRW
universe in which the Riemannian $3$-manifold $(\Sigma,\gamma)$ is
a compact Riemannian manifold of sufficiently small diameter, then
the horizon distance $d_{max}(t_0) = \int^{t_0}_{0} c/R(t) \, dt$
at the present time $t_0 \sim 10^{10} yrs$ may have exceeded the
diameter of $(\Sigma,\gamma)$, or may be a sufficient fraction of
the diameter that it is invalid to assume the observable spatial
universe has the topology of a solid ball $B^3$. Thus, even if one
were to accept that the observable spatial universe has almost no
spatial curvature, it would not follow that the observable spatial
universe has the topology of a solid ball.

Given $diam(\Sigma_t,\gamma_t) = R(t) \, diam(\Sigma,\gamma)$,
$\int^{t_0}_{0} c/R(t) \, dt \geq diam(\Sigma,\gamma)$ if and only
if $R(t_0) \int^{t_0}_{0} c/R(t) \, dt \geq
diam(\Sigma_{t_0},\gamma_{t_0}) $. If $d_{max}(t_0) \geq
diam(\Sigma,\gamma)$, the horizon would have disappeared, and we
would actually be able to see the entire spatial universe at the
present time. No point of the spatial universe could be separated
from us by a distance greater than $diam(\Sigma,\gamma)$, so if
$d_{max}(t_0) \geq diam(\Sigma,\gamma)$, then we would have
already received light from all parts of the spatial universe.
Individual galaxies and clusters of galaxies could produce
multiple images upon our celestial sphere without the occurrence
of gravitational lensing. Light emitted from opposite sides of a
galaxy could form images in opposite directions upon our celestial
sphere. Light emitted in different directions from a galaxy might
travel different distances before reaching us, and would therefore
produce images of different brightness. Furthermore, the light
which travelled the shorter distance would provide an image of the
galaxy as it appeared at a more recent stage of its evolution.
Light emitted by a galaxy in one direction could circumnavigate
the universe on multiple occasions and produce multiple `ghost
images' upon our celestial sphere. If $\Sigma$ were
non-orientable, light which had circumnavigated the universe an
odd number of times would produce a mirror image from light which
had circumnavigated the universe an even number of times.

One can define a compact FRW universe to be `small' if the size of
the horizon exceeds the size of the spatial universe,
$d_{max}(t_0) \geq diam(\Sigma,\gamma)$. In such universes, the
global spatial topology and geometry can have locally observable
consequences. Different compact spatial topologies and geometries
would produce different patterns of multiple and ghost images upon
the celestial sphere. In addition, a small compact universe would
produce patterns of paired circles in the Cosmic Microwave
Background Radiation (CMBR), (Cornish, Spergel and Starkman 1998).

Luminet and Roukema (1999, p15) point out that those compact
hyperbolic 3-manifolds with the smallest sizes, in curvature
radius units, are the most interesting in terms of cosmological
observational effects. This requires some explanation and
clarification. The size of a compact hyperbolic spatial universe
is not constrained by the topology of the compact hyperbolic
3-manifold. The size of a compact hyperbolic spatial universe can
vary arbitrarily. For a fixed compact hyperbolic 3-manifold, its
size must indeed be a fixed multiple of its `curvature radius',
(and powers thereof), but its curvature radius can vary
arbitrarily as a function of time. Moreover, there are two factors
to the curvature radius of the spatial universe: there is the
curvature radius $r$ used to define the scale factor on the
universal cover of the quotient construction that obtains the
spatial geometry $(\Sigma,\gamma)$; and there is the
time-dependent scale factor $R(t)$. Let $(\Sigma,\gamma) =
H^3(r)/\Gamma$. Then

$$
Vol \; (\Sigma,\gamma) = (Vol \; H^3/\Gamma)r^3
$$ and

$$\eqalign{
Vol \; (\Sigma_t,\gamma_t) &= (Vol \; (\Sigma,\gamma)) R(t)^3 \cr
&= (Vol \; H^3/\Gamma)r^3 R(t)^3 \, .}
$$ Hence, the size of a compact hyperbolic spatial universe is not a
fixed multiple of the time-dependent scale factor $R(t)$ and its
powers. Having fixed a compact hyperbolic topology for the spatial
universe, and having fixed a profile for the time-dependent scale
factor $R(t)$, one can independently vary the radius of curvature
$r$ in the universal cover $H^3(r)$ used to obtain that compact
hyperbolic topology. By so doing, one can arbitrarily vary the
volume of the spatial universe at the present time, $Vol \;
(\Sigma_{t_0},\gamma_{t_0})$, without changing either the topology
of the spatial universe or the profile of the time-dependent scale
factor $R(t)$. Whatever the compact hyperbolic topology chosen for
the spatial universe, one can judiciously choose a value of $r$
which enables the present spatial universe to fit inside the
present horizon. Conversely, whatever the compact hyperbolic
topology chosen for the spatial universe, one can always choose a
value of $r$ which makes the present spatial universe much larger
than the present horizon.

Suppose that one selects a compact hyperbolic topology which is
such that if one chooses $r=1$, then the current size of the
spatial universe will be much larger than the current size of the
particle horizon,

$$diam(\Sigma_{t_0},\gamma_{t_0})\gg R(t_0)\int^{t_0}_{0} \frac{c}{R(t)} \,
dt.$$ To remedy this situation, one can choose a very small value
for $r$ so that $1/r \gg 1$. Without changing the spatial
topology, this reduces the current size of the spatial universe to

$$
diam(\Sigma'_{t_0},\gamma'_{t_0}) = r \cdot
diam(\Sigma_{t_0},\gamma_{t_0}) \ll
diam(\Sigma_{t_0},\gamma_{t_0}) \,.
$$ Thus, a judicious choice of $r$ enables the present spatial universe
to fit inside the particle horizon.

However, varying $r$ does vary the current spatial curvature
$k(t_0) = k/R(t_0)^2$, which is constrained by observation. If one
begins with $r=1$ and $k=-1$, then re-setting $r \ll 1$ entails
changing to $k \ll -1$, and a change to a much more negative value
of $k(t_0) = k/R(t_0)^2$.

In most cosmology texts, the spatial curvature $k$ is set to $+1$,
$-1$ or $0$. Assuming $k \neq 0$, if one chooses a space form
$(\Sigma,\gamma)$ in which $|k| \neq 1$, i.e. if one chooses a
radius of curvature $r \neq 1$ on the universal cover of the space
form, then to re-set $k$ without changing the physical model, one
must re-set the time-dependent scale factor $R(t)$ so that it
incorporates the scale factor $r$. One re-sets the scale factor to

$$
R(t)_{new} = \frac{R(t)_{old}}{\sqrt{|k|}} = R(t)_{old}\, r \, .
$$ Note that $1/r = \sqrt{|k|}$.

Re-setting $k$ and $R(t)$ enables one to express the
time-dependence of the spatial curvature as

$$
k(t) = \frac{\pm 1}{R(t)_{new}^2} = \frac{\pm 1}{R(t)^2_{old}\cdot
r^2 }= \frac{k}{R(t)^2_{old}} \, .
$$

Assuming that $k \neq 0$, if observations suggest that the current
spatial curvature $k(t_0) = \pm 1/R(t_0)^2_{new}$ is very close to
zero, then it entails that $R(t_0)_{new}$ is very large. Hence,
unless the size of the compact hyperbolic manifold,
$diam(\Sigma,\gamma)$, is very small in curvature radius units,
then the current size of the spatial universe,
$diam(\Sigma_{t_0},\gamma_{t_0})$, will be much greater than the
current size of the particle horizon. Hence, those compact
hyperbolic 3-manifolds with the smallest sizes, in curvature
radius units, are the most interesting in terms of cosmological
observational effects.

\subsection{Inflation}

Cosmologists have postulated that the early universe underwent a
period of exponential, acceleratory expansion called `inflation'.
If inflation did take place, it means that the horizon distance
$d_{max}(t_0) = \int^{t_0}_{0} c/R(t) \, dt$ is much smaller than
it would have been otherwise. In the case of a compact spatial
universe $(\Sigma,\gamma)$, inflation makes the size of the
observable universe a smaller fraction of the size of the entire
universe than would otherwise have been the case. Given that
$R(t_0) \int^{t_0}_{0} c/R(t) \, dt$ is the present radius of the
observable spatial universe, and given that $R(t_0)\,
diam(\Sigma,\gamma)$ is the diameter of the present spatial
universe in the case of compact $(\Sigma,\gamma)$, the ratio

$$
\frac{R(t_0) \int^{t_0}_{0} c/R(t) \, dt}{R(t_0)\,
diam(\Sigma,\gamma)} = \frac{\int^{t_0}_{0} c/R(t) \,
dt}{diam(\Sigma,\gamma)}
$$ gives the present size of the observable spatial universe as a fraction
of the present size of the entire spatial universe. Clearly,
inflation reduces the value of the integral $\int^{t_0}_{0} c/R(t)
\, dt$, and therefore makes the size of the observable universe a
smaller fraction of the size of the entire universe than would
otherwise have been the case.

The advocates of inflation assert that in a universe which has
undergone inflation, the observable spatial universe must be very
much smaller than the entire spatial universe. This does not
follow from the last proposition, and is not necessarily the case.
If the present horizon distance $d_{max}(t_0) = \int^{t_0}_{0}
c/R(t) \, dt$ is reduced, it merely entails a decrease in the
diameter of the compact $3$-manifolds whose global topology and
geometry could have observable consequences at the present time.
Inflation does not entail that the observable spatial universe has
the topology of a solid ball.

It is widely asserted that if the early universe underwent a
period of inflationary expansion, which drove the time-dependent
scale factor to very high values, then the spatial curvature
$k/R(t_0)^2$ of the present universe must be very close to zero
even if the spatial universe is spherical or hyperbolic. Given
that
$$\Omega(t) = 1 + \frac{k}{H(t)^2 R(t)^2}\, ,$$ it is also
asserted that inflation produces a universe in which $\Omega$ is
very close to unity. These assertions rest upon the tacit
assumption that the sectional curvature $k$ of the Riemannian
manifold $(\Sigma,\gamma)$ in the warped product is very small. No
matter how large the time-dependent scale factor is, either as a
result of inflationary expansion or deceleratory FRW expansion,
the absolute value of $k$ can be chosen to be sufficiently large
that it cancels out the size of $R(t_0)^2$. With a judicious
choice of $k \gg 1$, a spherical universe which is 14 billion
years old, and which underwent inflation, could have spatial
curvature $k/R(t_0)^2 \approx 1$. Similarly, with a judicious
choice of $k \ll -1$, a hyperbolic universe which is 14 billion
years old, and which underwent inflation, could have spatial
curvature $k/R(t_0)^2 \approx -1$. (Whilst the issue is the
proximity of $k/R(t_0)^2$ to zero, I have chosen $| k/R(t_0)^2 |
\approx 1$ to represent a significant amount of spatial
curvature).

The time profile of the spatial curvature, $k(t) = k/R(t)^2 =
Sgn(k)/(R(t) \cdot r)^2 $, is determined by two independent
inputs: the sectional curvature $k$ of $(\Sigma,\gamma)$, and the
profile of the time-dependent scale factor, $R(t)$. The sectional
curvature $k$ is empirically meaningful: it is the value of $k(t)$
at the time that $R(t)=1$. If the profile of the time-dependent
scale factor, $R(t)$, is not fixed, then $k$ is not the value of
$k(t)$ at a fixed time in this family of models. The time at which
$R(t) =1$ varies from one model to another, depending upon the
functional expression chosen for $R(t)$. If $R(t)$ grows faster,
as it clearly does when $R(t)$ has the form of an exponential
function, then $R(t) =1$ at a much earlier time, and $k$ is the
value of $k(t)$ at a much earlier time. Having fixed a choice of
$R(t)$, one can independently vary $k$ to obtain an empirically
distinct family of models that share a time-dependent scale factor
with the same profile. Each such model has a different $k(t)$ time
profile. No matter how steep the chosen profile for $R(t)$, one
can find a family of models in which $k$ is sufficiently large
(or, equivalently, in which $r$ is sufficiently small), that the
present value of spatial curvature, $k(t_0) = k/R(t_0)^2 = \pm 1/
(R(t_0) \cdot r)^2$, remains significant.

For a spherical universe, whatever value inflation drives
$R(t_0)^2$ to, there is a value of sectional curvature
$0<k'<\infty$, such that for any $k \in [k',\infty)$, $k/R(t_0)^2
\geq 1$. There is only a finite range of values $(0,k')$ for which
$k/R(t_0)^2 < 1$, but an infinite range for which $k/R(t_0)^2 \geq
1$. Similarly, for a hyperbolic universe, there is a value of
sectional curvature $0>k''>-\infty$, such that for any $k \in
[k'',-\infty)$, $k/R(t_0)^2 \leq -1$. There is only a finite range
of values $(0,k'')$ for which $k/R(t_0)^2 > -1$, but an infinite
range for which $k/R(t_0)^2 \leq -1$. Whatever value inflation
drives $R(t_0)^2$ to, there is only a finite range of sectional
curvature values consistent with negligible spatial curvature in
the present day, but an infinite range which would produce a
significant amount of curvature in the present day.

Thus, inflation does not entail that a universe with an age of the
order $10^{10}yrs$ must have spatial curvature very close to zero.
This is true irrespective of whether the spatial universe is
compact or non-compact. Hence, contrary to the opinion held by
most cosmologists, the fact that our observable spatial universe
has a spatial curvature very close to zero, cannot be explained by
postulating a period of inflationary expansion alone. Even if the
entire spatial universe is very much larger than the observable
spatial universe, it does not entail that the observable spatial
universe must be approximately Euclidean. To explain the observed
`flatness' of our local spatial universe without preselecting
$\mathbb{R}^3$, one must either conjoin the postulate of inflation
with the postulate that $| k | \approx 1$, or the inflation-driven
growth in $R(t)$ must be commensurate with the magnitude of $k$.

Inflation is often presented as a solution to the flatness
`problem' and the horizon `problem', the latter of which will be
dealt with in the more extensive treatment of inflation contained
in part $\textrm{II}$ of this paper. The argument above is to the
effect that inflation does not solve the flatness `problem'.

\subsection{Dark energy}

Observations in the last decade using Type $\textrm{I}a$
supernovae as `standard candles' appear to indicate that the
expansion of our universe is accelerating. Type $\textrm{I}a$
supernovae at redshifts of $z \approx 0.5$ appear to be fainter
than would otherwise be the case. A universe which is currently
accelerating can be given an age estimate which is consistent with
the ages of the oldest stars in globular clusters. Under the
assumption of deceleratory expansion, the current value of the
Hubble parameter tended to yield an estimated age for the universe
which was less than the ages of the oldest stars in the universe.

The acceleratory expansion can be explained within general
relativity by an additional component to the energy density and
pressure in the Friedmann equations. This additional component,
referred to as `dark energy', must be such that it produces a
repulsive gravitational effect. The structure of a Lorentzian
manifold itself is not sufficient to guarantee that gravity will
be attractive. Given a timelike vector $Z \in T_p \mathcal{M}$,
one has a tidal force operator $F_Z: Z^\bot \rightarrow Z^\bot$
defined on each $W \in Z^\bot$ by

$$F_Z(W) = R(\,\cdot \,,Z,W,Z)\, ,$$ where $R$ is the Riemann curvature tensor.
$F_Z(W)$ is physically interpreted as the tidal force due to
gravity, acting on a particle in spatial direction $W$, for an
instantaneous observer $Z$. Hence, $\langle F_Z(W), W \rangle \leq
0$ means that gravity is attractive in direction $W$. Now, the
trace of the tidal force operator,

$$
\text{tr}\; F_Z = \sum_{i=1}^{3} \langle F_Z(e_i), e_i \rangle =
-Ric(Z,Z)\, ,
$$ where $\{e_1,e_2,e_3\}$ is an orthonormal basis of the subspace
$Z^\bot$, gives the sum of the tidal force over three orthogonal
directions in $Z^\bot$. Hence, the `timelike convergence
condition' that $Ric(Z,Z) \geq 0$, for all timelike vectors $Z$,
stipulates that the net effect of gravity is attractive.

A component in the Friedmann equations will have a repulsive
gravitational effect if it possesses a negative pressure $p <
-\frac{1}{3}\rho$. Thus, in terms of an `equation of state', which
expresses pressure as a function of energy density $p = f(\rho)$,
dark energy is such that $p = w\rho$ with $w< - \frac{1}{3}$.

A non-zero and positive cosmological constant provides a special
case of dark energy. In the general case of a non-zero
cosmological constant $\Lambda$, the Einstein field equations
become

$$
8\pi G \, T_{\mu \nu} - \Lambda g_{\mu\nu} = R_{\mu\nu} - 1/2 \;
\text{S} \, g_{\mu\nu}\, ,
$$ and the Friedmann equations become, (Heller 1992,
p101):

$$
\frac{8\pi G}{3}\rho(t) = \left( \frac{R'(t)}{R(t)} \right)^2 +
\frac{k}{R(t)^2} - \frac{1}{3}\Lambda \, ,
$$

$$
-8\pi G p(t) = 2 \frac{R''(t)}{R(t)} + \left( \frac{R'(t)}{R(t)}
\right) ^2 + \frac{k}{R(t)^2} - \Lambda \,.
$$

The presence of a non-zero cosmological constant is equivalent to
an additional component of the energy density and pressure in the
Friedmann equations without cosmological constant. Let $\rho_m$
denote the energy density due to matter alone. With the
cosmological constant, the Friedmann equation for energy density
can be written as

$$
\frac{8\pi G}{3}\rho_m(t) + \frac{1}{3}\Lambda = \left(
\frac{R'(t)}{R(t)} \right)^2 + \frac{k}{R(t)^2} \,.
$$ Setting $\rho_\Lambda = \Lambda/8\pi G$
one can further re-write this equation as

$$
\frac{8\pi G}{3}(\rho_m(t) + \rho_\Lambda) = \left(
\frac{R'(t)}{R(t)} \right)^2 + \frac{k}{R(t)^2} \,.
$$ Hence, the presence of a non-zero cosmological constant corresponds to
an additional, \emph{time-independent} component $\rho_\Lambda$ to
the energy density. One re-defines $\Omega$ as

$$
\Omega = \frac{\rho_m + \rho_\Lambda}{\rho_c} = \frac{\rho_m +
\rho_\Lambda}{3H^2/8\pi G} = \frac{\rho_m}{3H^2/8\pi
G}+\frac{\rho_\Lambda}{3H^2/8\pi G} \equiv \Omega_m +
\Omega_\Lambda \,.
$$

With the cosmological constant, the Friedmann equation for
pressure can be written as

$$
-8\pi G p(t) + \Lambda = 2 \frac{R''(t)}{R(t)} + \left(
\frac{R'(t)}{R(t)} \right) ^2 + \frac{k}{R(t)^2} \,.
$$ Setting $p_\Lambda = -\Lambda/8\pi G$, so $\Lambda = - p_\Lambda 8\pi G$,
one can further re-write this equation as

$$
-8\pi G (p(t) - p_\Lambda)  = 2 \frac{R''(t)}{R(t)} + \left(
\frac{R'(t)}{R(t)} \right) ^2 + \frac{k}{R(t)^2} \,.
$$

A positive cosmological constant behaves like an repulsive
component to gravity because $p_\Lambda = -\rho_\Lambda$. In terms
of an equation of state $p_\Lambda = w\rho_\Lambda$, a
cosmological constant has $w=-1$.

In the case of a non-zero cosmological constant, the dark energy
can be interpreted as a property of space-time, rather than a
property of some exotic field in space-time. If the additional
component to the energy density in the Friedmann equations without
cosmological constant represents some exotic field, then this
additional energy density can be time-dependent and can possess a
time-dependent equation of state. In contrast, a non-zero
cosmological constant can only correspond to a constant energy
density and a constant equation of state. The latest astronomical
evidence, (Adam G.Riess \textit{et al} 2004), indicates that the
dark energy is a non-zero cosmological constant.

If the cosmological constant is non-zero, one can no longer infer
the sign of the spatial curvature and the long-term dynamical
behaviour of a FRW universe from $\Omega_m$, but because the
Friedmann equation is unchanged when the cosmological constant is
incorporated into the total $\rho$, they can still be inferred
from $\Omega$, with $\Omega = \Omega_m + \Omega_\Lambda$ in this
case.

The current observational evidence leads cosmologists to believe
that $\Omega$ is approximately unity, with $\Omega_m \approx 0.3$
and $\Omega_\Lambda \approx 0.7$. It might, however, be noted that
circa 1967, observations appeared to indicate a surplus number of
quasars at redshift $z=2$, (Heller 1992, p102). These observations
were explained by postulating a Lemaitre model, a FRW model with a
positive cosmological constant, in which the expansion of the
universe is punctuated by an almost static period, a type of
plateau in the scale factor when it is displayed as a function of
time. The quasar observations transpired to be a selection effect,
and the community of cosmologists reverted to their belief that
$\Lambda = 0$. Perhaps in a similar vein, it has been suggested
that the dimming of Type $\textrm{I}a$ supernovae at redshifts of
$z \approx 0.5$ could be due to screening from `grey dust', or due
to intrinsically fainter Type $\textrm{I}a$ supernovae at
redshifts of $z \approx 0.5$.

\section{Spatially homogeneous cosmologies}

As a continuation to the rationale of the opening section, the
philosophical purpose of this section is to explain and emphasise
the immense variety of spatially homogeneous cosmological models
which are consistent with astronomical observation, or which serve
to highlight the variety of possible universes similar to our own.
This section will also clarify the Bianchi classification, and the
relationship between the spatially homogeneous models and the FRW
models.

\hfill \break

The spatially homogeneous class of cosmological models are usually
presented as a generalisation of the Friedmann-Robertson-Walker
cosmological models. The generalisation is said to be obtained by
dropping the requirement of spatial isotropy, but retaining the
requirement of spatial homogeneity. The FRW models are considered
to be special cases of the class of spatially homogeneous models.

The topology of a typical spatially homogeneous cosmological model
is a product $I \times \Sigma$ of an open interval $I \subset
\mathbb{R}^1$ with a connected $3$-dimensional manifold $\Sigma$.
The $4$-dimensional manifold $\mathcal{M} = I \times \Sigma$ is
ascribed a Lorentzian metric tensor which induces a homogeneous
Riemannian metric $\gamma_t$ on each hypersurface $\Sigma_t = t
\times \Sigma$. Thus, each pair $(\Sigma_t,\gamma_t)$ is a
homogeneous $3$-dimensional Riemannian manifold. A spatially
homogeneous cosmological model is a Lorentzian manifold
$\mathcal{M}$ in which the orbits of the isometry group
$I(\mathcal{M})$ consist of such a one-parameter family of
spacelike hypersurfaces.

As with the FRW models, there is a spatial topology $\Sigma$
associated with each spatially homogeneous cosmological model.
However, the spatial geometry of a spatially homogeneous model can
vary in a more complex manner than the single scale factor
variation of a FRW model. In other words, there is no need for a
spatially homogeneous model to be a warped product. The class of
cosmological models obtained by taking warped products $I \times
_R \Sigma$ in which $\Sigma$ is a (globally) homogeneous
$3$-dimensional Riemannian manifold, only constitutes a proper
subset of the entire class of spatially homogeneous cosmological
models.

The time variation of the spatial geometry in a spatially
homogeneous cosmology is, in general, expressed by a matrix of
scale factors, rather than a single scale factor. Whilst a warped
product geometry can be expressed as $-dt \otimes dt + R(t)^2
\gamma$, in a general spatially homogeneous cosmology, each
component of spatial geometry can be subject to time variation,
hence the metric can be expressed as $-dt \otimes dt +
\gamma_{ab}(t)\omega^a(t) \otimes \omega^b(t)$, where
$\omega^a(t), a=1,2,3$ are one-forms on $\Sigma_t$ invariant under
the action of the isometry group $I(\Sigma_t)$.

Now consider a connected $3$-dimensional homogeneous Riemannian
manifold $(\Sigma,\gamma)$. Associated with $(\Sigma,\gamma)$ are
the isometry group $I(\Sigma)$ and the isotropy subgroup $H$. The
isometry group can be of dimension 6,4, or 3. Moreover, it is true
that

$$
\text{dim}\; \Sigma = \text{dim}\; I(\Sigma) - \text{dim} \; H \,.
$$ Thus, when $I(\Sigma)$ is of dimension 6, $H$ will be of
dimension 3; when $I(\Sigma)$ is of dimension 4, $H$ will be of
dimension 1; and when $I(\Sigma)$ is of dimension 3, the isotropy
group $H$ will be trivial.

In any dimension, it can be shown that every homogeneous
Riemannian manifold $(\Sigma, \gamma)$ is diffeomorphic to some
Lie group. In particular, every homogeneous $3$-dimensional
Riemannian manifold $(\Sigma, \gamma)$ is diffeomorphic to some
$3$-dimensional Lie group. The $3$-dimensional Riemannian manifold
$\Sigma$ is diffeomorphic with the quotient Lie group
$I(\Sigma)/H$, the quotient of the isometry group by the isotropy
subgroup.

If one has a homogeneous $3$-dimensional Riemannian manifold
$(\Sigma,\gamma)$ which has a 3-dimensional isometry group
$I(\Sigma)$, then $I(\Sigma)/H \cong I(\Sigma)$, and the
Riemannian manifold is diffeomorphic with its own isometry group.

In the event that a homogeneous $3$-dimensional Riemannian
manifold $(\Sigma,\gamma)$ has an isometry group $I(\Sigma)$ of
dimension 4 or 6, the quotient $I(\Sigma)/H$ will be distinct from
$I(\Sigma)$, and the Riemannian manifold will not be diffeomorphic
with its own isometry group.

By the definition of homogeneity, the isometry group $I(\Sigma)$
of a homogeneous Riemannian manifold $(\Sigma,\gamma)$ must act
transitively. However, when the isometry group $I(\Sigma)$ is of
dimension 3, the action is simply transitive, and when $I(\Sigma)$
is of dimension 4 or 6, the action is multiply transitive.

Not only is every homogeneous Riemannian manifold $(\Sigma,
\gamma)$ diffeomorphic to some Lie group, but conversely, any Lie
group can be equipped with a metric which renders it a homogeneous
Riemannian manifold. Thus, the topologies of all the
$3$-dimensional Lie groups equal the possible topologies for a
$3$-dimensional homogeneous Riemannian manifold. A list of all the
$3$-dimensional Lie groups will exhaust the possible topologies
for a $3$-dimensional homogeneous Riemannian manifold. However,
this list of topologies is repetitious; although every
$3$-dimensional Lie group will provide the topology for a
$3$-dimensional homogeneous Riemannian manifold, two distinct Lie
groups can possess the same topology.

To obtain a classification of all the connected $3$-dimensional
Lie groups, the first step is to obtain a classification of all
the simply connected $3$-dimensional Lie groups. Simply connected
Lie groups are in a one-to-one correspondence with Lie algebras,
and there is a classification of the isomorphism classes of
$3$-dimensional Lie algebras called the Bianchi classification.
Hence, the Bianchi classification provides a classification of the
simply connected $3$-dimensional Lie groups. The Bianchi
classification of all the $3$-dimensional Lie algebras only
provides a coarse-grained classification of the connected
$3$-dimensional Lie groups because many Lie groups can possess the
same Lie algebra. However, all the Lie groups which share the same
Lie algebra will be `locally isomorphic', and will have a common
simply connected, universal covering Lie group. Each connected
$3$-dimensional Lie group $G$ is obtained from its universal cover
$\widetilde{G}$ as the quotient $\widetilde{G}/N$ of its universal
cover with respect to a discrete, normal subgroup $N$. If
$\widetilde{G}$ has Lie algebra $\mathfrak{g}$, then the quotient
$\widetilde{G}/N$ will also have Lie algebra $\mathfrak{g}$. A
discrete normal subgroup of a connected Lie group is contained in
the centre of the Lie group, hence $N$ is a central, discrete,
normal subgroup.

Once the simply connected $3$-dimensional Lie groups have been
classified, the second step is to classify all the discrete normal
subgroups of each simply connected $3$-dimensional Lie group, up
to conjugacy. Step two yields a family of Lie groups
$\widetilde{G}/N_i,\widetilde{G}/N_j,...$ which share the same
Bianchi type, but which are distinct, possibly non-diffeomorphic
Lie groups. These two steps together provide a classification of
all the connected $3$-dimensional Lie groups.

To reiterate, a list of all the connected $3$-dimensional Lie
groups provides an exhaustive, but repetitious list of all the
possible homogeneous spatial topologies. Note that a list which
exhausts all the possible homogeneous spatial topologies, does not
provide a list of all the possible homogeneous spatial geometries.
A $3$-dimensional Lie group can support more than one homogeneous
metric.

\hfill \break

Let us turn, then, to the Bianchi classification of the
isomorphism classes of $3$-dimensional Lie algebras. Given a
choice of basis $\{e_1,e_2,e_3\}$ for a $3$-dimensional Lie
algebra, the structure constants $C_{ij}^k$ are defined to be such
that $[e_i,e_j] = C_{ij}^k e_k$. The Bianchi classification is
based upon the fact that Lie algebras can be characterised in
terms of their structure constants $C_{ij}^k$, and the fact that
for a $3$-dimensional Lie algebra, the structure constants can be
expressed as

$$
C_{ij}^k = \epsilon_{ijl}B^{lk} + \delta^k_j a_i - \delta^k_i a_j
\, ,
$$ where $B$ is a symmetric $3 \times 3$ matrix, and $\textbf{a}$ is a $1
\times 3$ column vector, (Dubrovin \textit{et al} 1992, Part I,
\S24.5, p230). The Jacobi identity which constrains the structure
constants of a Lie algebra entails that

$$
B^{ij}a_j = 0 \,.
$$

Although the structure constants of a Lie algebra are
basis-dependent, the classification of $3$-dimensional Lie
algebras is basis-independent. Hence, the classification uses the
fact that one can choose a basis in which $B$ is a diagonal matrix
with $B^{ii} = \pm 1, 0$ for $i =1,2,3$, and $\textbf{a} =
(a,0,0)$. In this basis, the structure constants are such that

$$\eqalign{
[e_1,e_2] &= a e_2 + B^{33}e_3 \cr [e_2,e_3] &= B^{11}e_1 \cr
[e_3,e_1] &= B^{22}e_2 - a e_3 \,.}
$$

With this choice of basis, it also follows that $B^{11}a = 0$,
hence either $B^{11}$ or $a$ is zero. The Bianchi types, denoted
by Roman numerals, are duly defined in Table \ref{Bianchi}.

\begin{table}[h]
\caption{Bianchi classification of 3-dimensional Lie algebras}
\label{Bianchi}
\begin{center}
\begin{tabular}{|c|c|c|c|c|}
  \noalign{\hrule}
Type & a & $B^{11}$ & $B^{22}$ & $B^{33}$ \\ \hline
  $\textrm{I}$ & 0 & 0 & 0 & 0 \\
  $\textrm{II}$ & 0 & 1 & 0 & 0 \\
$\textrm{VI}_0$ & 0 & 1 & -1 & 0 \\
$\textrm{VII}_0$ & 0 & 1 & 1 & 0 \\
$\textrm{VIII}$ & 0 & 1 & 1 & -1 \\
 $\textrm{IX}$ & 0 & 1 & 1 & 1 \\
  $\textrm{V}$ & 1 & 0 & 0 & 0 \\
 $\textrm{IV}$ & 1 & 0 & 0 & 1 \\
$\textrm{VI}_h \; (h < 0) \; a \neq 1$ & $\sqrt{-h}$ & 0 & 1 & -1 \\
 $\textrm{III}$ & 1 & 0 & 1 & -1 \\
 $\textrm{VII}_h \; (h > 0)$ & $\sqrt{h}$ & 0 & 1 & 1 \\ \hline
\end{tabular}
\end{center}
\end{table}

Note that $\textrm{III} = \textrm{VI}_{-1}$ if we remove the
restriction that $a \neq 1$ for type $\textrm{VI}_h$.

\hfill \break

Before proceeding further, some salient definitions concerning Lie
algebras are required. Given a Lie algebra $\mathfrak{g}$, one can
inductively define the \emph{lower central series} of subalgebras
$\mathscr{D}_k\mathfrak{g}$ by

$$
\mathscr{D}_1\mathfrak{g} = [\mathfrak{g},\mathfrak{g}], \; \;
\mathscr{D}_k\mathfrak{g} = [\mathfrak{g},
\mathscr{D}_{k-1}\mathfrak{g}] \,.
$$ A Lie algebra is defined to be nilpotent if $\mathscr{D}_k\mathfrak{g}=
0$ for some $k$.

Secondly, one can inductively define the \emph{derived series} of
subalgebras $\mathscr{D}^k\mathfrak{g}$ by

$$
\mathscr{D}^1\mathfrak{g} = [\mathfrak{g},\mathfrak{g}], \; \;
\mathscr{D}^k\mathfrak{g} = [\mathscr{D}^{k-1}\mathfrak{g},
\mathscr{D}^{k-1}\mathfrak{g}] \,.
$$ A Lie algebra is defined to be solvable if $\mathscr{D}^k\mathfrak{g}=
0$ for some $k$.

An ideal in a Lie algebra is a Lie subalgebra $\mathfrak{h}
\subset \mathfrak{g}$ which is such that $[X , Y ] \in
\mathfrak{h}$ for all $X \in \mathfrak{h}$, $Y \in \mathfrak{g}$.
A Lie algebra can be defined to be semi-simple if it has no
nonzero solvable ideals. (Fulton and Harris 1991, p122-123).

Under the Bianchi classification, there are six `Type A' Lie
algebras: $\textrm{I}$, $\textrm{II}$, $\textrm{VI}_0$,
$\textrm{VII}_0$, $\textrm{VIII}$, and $\textrm{IX}$. These are
the Lie algebras of the six unimodular $3$-dimensional connected
Lie groups. As Lie algebras, they are trace-free. All the other
Lie algebras are `Type B'.

Bianchi types $\textrm{VIII}$ and $\textrm{IX}$ are the only
semi-simple real $3$-dimensional Lie algebras. All the other
Bianchi types are solvable Lie algebras.\footnote{The ensuing
discussion of simple, solvable, and nilpotent Lie algebras was
motivated by a private communication from Karl H.Hofmann} In
particular, Bianchi types $\textrm{VIII}$ and $\textrm{IX}$ are
both simple Lie algebras. Type $\textrm{VIII}$ is
$\mathfrak{sl}(2,\mathbb{R}) \cong \mathfrak{so}(2,1)$, and type
$\textrm{IX}$ is $\mathfrak{so}(3) \cong \mathfrak{su}(2)$.

Bianchi types $\textrm{I}$ and $\textrm{II}$ are the only
nilpotent solvable real $3$-dimensional Lie algebras. The Bianchi
type $\textrm{I}$ is the abelian Lie algebra $\mathbb{R}^3$, and
any abelian Lie algebra is automatically nilpotent and solvable.
The Bianchi type $\textrm{II}$ is the $3$-dimensional Heisenberg
Lie algebra, a $2$-step nilpotent Lie algebra,

$$
[\mathfrak{g},\mathfrak{g}] \neq 0, \; \;
[\mathfrak{g},[\mathfrak{g},\mathfrak{g}]] = 0 \, ,
$$ with a $1$-dimensional centre.

The other seven classes of Lie algebra all contain non-nilpotent
solvable Lie algebras. Of the `Type A' Lie algebras,
$\textrm{VI}_0$ and $\textrm{VII}_0$ are the non-nilpotent
solvable ones. Type $\textrm{VI}_0$ is the Lie algebra of
$E(1,1)$, the group of motions of the Euclidean plane equipped
with a Minkowski metric. Type $\textrm{VII}_0$ is the Lie algebra
of $E(2)$, the group of motions of the Euclidean plane equipped
with a spacelike metric.

Within the `Type B' Lie algebras, Bianchi types $\textrm{VI}_h$
and $\textrm{VII}_h$ provide one-parameter families of Lie
algebras for $0 < h < \infty$, for which the $\textrm{VI}_0$ and
$\textrm{VII}_0$ Lie algebras are limiting cases as $h \rightarrow
0$.

The Type B Bianchi algebra $\textrm{III}$ is such that
$\textrm{III} = \textrm{VI}_{-1}$ if we remove the restriction
that $a \neq 1$ for type $\textrm{VI}_h$. The Bianchi type
$\textrm{III}$ algebra brings us to the Levi-Malcev decomposition.

The sum of all the solvable ideals in a Lie algebra $\mathfrak{g}$
is a maximal solvable ideal called the radical $\mathfrak{r}$. The
quotient $\mathfrak{g}/\mathfrak{r}$ is a semi-simple Lie algebra.
There exist mutually conjugate subalgebras of $\mathfrak{g}$,
called Levi subalgebras $\mathfrak{l}$, which are maximal
semi-simple subalgebras, and which map isomorphically onto
$\mathfrak{g}/\mathfrak{r}$. The Levi-Malcev decomposition states
that for any Lie algebra $\mathfrak{g}$, there is a Levi
subalgebra $\mathfrak{l}$ such that $\mathfrak{g} =\mathfrak{r}
\oplus \mathfrak{l}$.

Now, a semi-simple Lie algebra has no non-zero solvable ideals,
hence a semi-simple algebra has no radical. A real $3$-dimensional
semi-simple Lie algebra therefore has a trivial Levi
decomposition, coinciding with its own Levi subalgebra. On the
other hand, a solvable Lie algebra is coincident with its own
radical, so it too has a trivial Levi decomposition. A real
$3$-dimensional Lie algebra with a non-trivial Levi decomposition
would have to be the sum of one $2$-dimensional Lie algebra and a
$1$-dimensional Lie algebra. Now, there is only one real
$1$-dimensional Lie algebra, the abelian Lie algebra $\mathbb{R}$.
Being abelian, it must be solvable and not semi-simple, hence it
could only provide the radical $\mathfrak{r}$ in the Levi
decomposition of a real $3$-dimensional Lie algebra. The Levi
subalgebra $\mathfrak{l}$ in such a decomposition would therefore
have to be a real $2$-dimensional semi-simple Lie algebra. In
fact, a real semi-simple Lie algebra is at least
$3$-dimensional,\footnote{Private communication with Karl
H.Hofmann} hence no real $3$-dimensional Lie algebra possesses a
non-trivial Levi decomposition. To reiterate, every real
$3$-dimensional Lie algebra is either semi-simple or solvable.
There is a unique non-abelian real $2$-dimensional Lie algebra,
$V^2$, but it is not semi-simple. The type $\textrm{III}$ Bianchi
algebra is the direct sum $\mathbb{R} \oplus V^2$, but the type
$\textrm{III}$ algebra is solvable, and this is not a Levi
decomposition.

\hfill \break

The spatially homogeneous cosmological models in which each pair
$(\Sigma_t,\gamma_t)$ has a $3$-dimensional isometry group are
referred to as Bianchi cosmological models. The case in which
$I(\Sigma_t)$ is $3$-dimensional is obviously the case in which
the isotropy group at each point is trivial. In this case, the
Riemannian manifold $\Sigma_t$ is diffeomorphic to the isometry
group $I(\Sigma_t)$. Moreover, in this case, the Bianchi
classification of $3$-dimensional Lie algebras can contribute to
the classification of the homogeneous spatial geometries because
the Lie algebra of Killing vector fields on $\Sigma_t$ is
isomorphic with the Lie algebra of the isometry group
$I(\Sigma_t)$. Those homogeneous $3$-dimensional Riemannian
manifolds which have $3$-dimensional isometry groups can be
classified according to Bianchi types \textrm{I} - \textrm{IX}. To
be clear, the Lie algebra of Killing vector fields on a
homogeneous $3$-dimensional Riemannian manifold $\Sigma_t$ is
always isomorphic with the Lie algebra of the isometry group
$I(\Sigma_t)$, but the Bianchi classification only provides a
classification of the Lie algebras of Killing vector fields in the
case in which $I(\Sigma_t)$ is $3$-dimensional.

Groups with the same Lie algebra are not, in general, isomorphic
Lie groups, hence $3$-dimensional homogeneous geometries of the
same Bianchi type do not, in general, have the same
$3$-dimensional isometry groups, and are not, in general,
isometric geometries. Distinct geometries can share the same Lie
algebra of Killing vector fields.

The spatially homogeneous models in which each pair
$(\Sigma_t,\gamma_t)$ has a 4-dimensional isometry group are
called rotationally symmetric in contrast with the spherical
symmetry of the FRW models. Whilst the isotropy group at each
point of an FRW model contains the $3$-dimensional group $SO(3)$,
the isotropy group at each point of a rotationally symmetric model
is the $1$-dimensional group $SO(2)$. This $1$-dimensional group
acts transitively upon the set of directions within a
$2$-dimensional plane of the tangent space. The rotationally
symmetric models are often referred to as Kantowski-Sachs models.
In fact, the latter term should be reserved for models with a
$4$-dimensional isometry group $I(\Sigma_t)$ in which there is no
$3$-dimensional isometry subgroup which acts simply transitively
upon $\Sigma_t$.

Finally, those spatially homogeneous models in which each pair
$(\Sigma_t,\gamma_t)$ has a $6$-dimensional isometry group, and in
which the time variation of the spatial geometry is given by a
single scale factor, are Friedmann-Robertson-Walker models. It is
true that the conventional FRW models, which are spatially
globally isotropic, are special cases of the class of spatially
homogeneous cosmological models. A globally isotropic
$3$-dimensional Riemannian manifold does indeed have a
6-dimensional isometry group. However, the generalized class of
FRW models takes one outside the class of spatially homogeneous
models. A locally isotropic $3$-dimensional Riemannian manifold
need not be homogeneous. It is therefore incorrect to consider the
entire class of FRW models as a subclass of the spatially
homogeneous cosmological models. In many of the generalized FRW
models, the spatial geometry has an isometry group of dimension
lower than 6, or no Lie group of isometries at all.

Although the 6-dimensional isometry group $I(\Sigma_t)$ of each
hypersurface $\Sigma_t$ in a spatially homogeneous
Friedmann-Robertson-Walker model is not diffeomorphic with the
Riemannian manifold $\Sigma_t$, it does contain $3$-dimensional
Lie subgroups which act simply transitively upon $\Sigma_t$, and
the Lie algebras of these subgroups do fall under the Bianchi
classification. The isometry group of the $\mathbb{R}^3$ FRW model
contains the Bianchi type $\textrm{I}$ group of translations on
$\mathbb{R}^3$ as a simply transitive subgroup. The isotropy group
at each point is a $3$-dimensional subgroup of Bianchi type
$\textrm{VII}_0$. In the case of the $H^3$ FRW model, the isometry
group contains a simply transitive $3$-dimensional subgroup of
Bianchi type $\textrm{V}$, whilst the isotropy group is a
$3$-dimensional subgroup of type $\textrm{VII}_h$. In the case of
the $S^3$ FRW model, the isometry group contains a simply
transitive $3$-dimensional subgroup of Bianchi type $\textrm{IX}$,
whilst the isotropy group is a $3$-dimensional subgroup also of
type $\textrm{IX}$. (Rey and Luminet 2003, p43-44).

In contrast, the spatially homogeneous Kasner cosmology has a
$3$-dimensional isometry group of Bianchi type $\textrm{I}$, and
no isotropy group, whilst the spatially homogeneous Mixmaster
cosmology has a $3$-dimensional isometry group of Bianchi type
$\textrm{IX}$, and no isotropy group.

Note that in space-times which can be sliced up into a
one-parameter family of homogeneous spacelike hypersurfaces
$(\Sigma_t,\gamma_t)$, each bearing a specific Bianchi type, there
is no guarantee that the Bianchi type of each homogeneous
hypersurface will be the same; the Bianchi type can change with
time, (Rainer and Schmidt 1995).

A $3$-dimensional geometry whose isometry group admits a simply
transitive $3$-dimensional subgroup, must be homeomorphic with
that $3$-dimensional group. All the simply connected
$3$-dimensional Lie groups in the Bianchi classification are
either diffeomorphic to $\mathbb{R}^3$, or diffeomorphic to $S^3$
in the case of the $\textrm{IX}$ type. Hence, any globally
homogeneous $3$-dimensional geometry which falls under the Bianchi
classification, has a topology which is either covered by
$\mathbb{R}^3$ or $S^3$.

In the case of a homogeneous $3$-dimensional Riemannian manifold
$\Sigma$, which has a $3$-dimensional isometry group $I(\Sigma)$
to which it is diffeomorphic, a quotient $\Sigma/\Gamma$ with
respect to a discrete normal subgroup of $I(\Sigma)$ is
diffeomorphic to the quotient Lie group $I(\Sigma)/\Gamma$. Given
that the isometry group of such a quotient is $N(\Gamma)/\Gamma$,
and given that $\Gamma$ is a normal subgroup, $N(\Gamma) =
I(\Sigma)$, and it follows that the isometry group of the quotient
is $I(\Sigma)/\Gamma$. If the manifold is diffeomorphic with its
isometry group, then the quotient manifold is diffeomorphic with
the isometry group of the quotient. Given that the isometry group
of the quotient can also be expressed as the centralizer
$Z(\Gamma)$, it follows that the quotient is a homogeneous
Riemannian manifold if and only if the centralizer $Z(\Gamma)$
acts transitively on $\Sigma$, (Ellis 1971, p11). Given that a
discrete normal subgroup of a connected Lie group must be central,
the centralizer $Z(\Gamma)$ will, in this instance, contain
$\Gamma$ as a subgroup.

\hfill \break

Assuming the Copernican principle is true, the observed local
isotropy of our universe can be used to exclude a number of
$3$-manifolds which would otherwise be candidates for the spatial
topology. The reasoning here follows from two key facts:

\begin{enumerate}
\item A locally isotropic $3$-dimensional Riemannian manifold must
be of constant sectional curvature (Wolf 1967, p381-382)
\item A $3$-dimensional manifold can only possess a Riemannian
metric tensor of constant sectional curvature if its universal
covering manifold is diffeomorphic to either $\mathbb{R}^3$ or
$S^3$.
\end{enumerate}

A simply connected manifold is its own universal cover, hence a
simply connected $3$-manifold which is not diffeomorphic to either
$\mathbb{R}^3$ or $S^3$ will not be able to support a Riemannian
metric tensor of constant sectional curvature, and will therefore
not be able to represent a locally isotropic spatial universe.

To find such a manifold, we note that for a connected product
manifold $M \times N$, the fundamental group $\pi_1(M \times N)$
is such that

$$
\pi_1(M \times N) \cong \pi_1(M) \times \pi_1(N) \,.
$$ Hence, if $M$ and $N$ are both simply connected, then $M \times
N$ will be simply connected. $S^2$ and $\mathbb{R}^1$ are both
simply connected, hence the hypercylinder $S^2 \times
\mathbb{R}^1$ is also simply connected.

$S^2 \times \mathbb{R}^1$ is not diffeomorphic to either
$\mathbb{R}^3$ or $S^3$, hence the hypercylinder $S^2 \times
\mathbb{R}^1$ cannot support a Riemannian metric tensor of
constant sectional curvature, and cannot therefore represent a
locally isotropic spatial universe.

Including $S^2 \times \mathbb{R}^1$ itself, there are seven
$3$-manifolds which have $S^2 \times \mathbb{R}^1$ as their
universal covering, (Scott 1983, p457-459). Each such manifold has
a universal covering which is neither $\mathbb{R}^3$ nor $S^3$,
hence each such manifold cannot support a Riemannian metric tensor
of constant sectional curvature, and cannot therefore represent a
locally isotropic spatial universe. Of these seven manifolds,
three are non-compact and four are compact. The non-compact cases
consist of $S^2 \times \mathbb{R}^1$ itself, the trivial line
bundle $\mathbb{RP}^2 \times \mathbb{R}^1$, and a non-trivial line
bundle over $\mathbb{RP}^2$. The compact cases consist of
$\mathbb{RP}^2 \times S^1$, the connected sum $\mathbb{RP}^3 \#
\mathbb{RP}^3$, and a pair of line bundles over $S^2$, one of
which is the trivial bundle $S^2 \times S^1$.

\hfill \break

Thurston has identified eight globally homogeneous, simply
connected $3$-dimensional Riemannian manifolds which admit a
compact quotient, (Rey and Luminet 2003, p39-42). $\mathbb{R}^3$,
$S^3$, and $H^3$ provide three of these, but the remaining five
are non-isotropic. These five geometries and their quotients are
neither globally nor locally isotropic. Moreover, the quotients of
these five geometries are only guaranteed to be locally
homogeneous. The hypercylinder $S^2 \times \mathbb{R}^1$ is
obviously one of these five geometries. The others are $H^2 \times
\mathbb{R}^1$, $\widetilde{SL2\mathbb{R}}$, the universal covering
of the $3$-dimensional group $SL(2,\mathbb{R})$, $Nil$, the
$3$-dimensional Lie group of $3 \times 3$ Heisenberg matrices, and
$Sol$, a Lie group consisting of $\mathbb{R}^3$ equipped with a
non-standard group product.

The quotients of $H^2 \times \mathbb{R}^1$ include all the
products of $T_g$ with either $S^1$ or $\mathbb{R}^1$, where the
$T_g$ are the compact, orientable surfaces of genus $g > 1$,
equipped with metrics of constant negative curvature, and
constructed from the $2$-sphere by attaching $g$ handles.

$Sol$ has a disconnected isometry group with eight components, the
identity component of which is $Sol$ itself, (Koike \textit{et al}
1994, p12). The other four geometries possess a $4$-dimensional
isometry group. $S^2 \times \mathbb{R}^1$, $H^2 \times
\mathbb{R}^1$, $\widetilde{SL2\mathbb{R}}$ and $Nil$ are therefore
rotationally symmetric models. However, only the hypercylinder
$S^2 \times \mathbb{R}^1$ provides a Kantowski-Sachs model. Whilst
the isometry group of $S^2 \times \mathbb{R}^1$ has no
$3$-dimensional subgroups which act simply transitively upon $S^2
\times \mathbb{R}^1$, the isometry group of $H^2 \times
\mathbb{R}^1$ has a Bianchi type $\textrm{III} = \textrm{VI}_{-1}$
subgroup, the isometry group of $\widetilde{SL2\mathbb{R}}$ has a
Bianchi type $\textrm{VIII}$ subgroup, and the isometry group of
$Nil$ has a Bianchi type $\textrm{II}$ subgroup, each of which
acts simply transitively. The isometry group of $Sol$ is a Bianchi
type $\textrm{VI}_0$ group. (Rey and Luminet 2003, p45).

There are only three distinct topologies amongst the eight
Thurston geometries. $\mathbb{R}^3$, $H^3$, $H^2 \times
\mathbb{R}^1$, $\widetilde{SL2\mathbb{R}}$, $Nil$, and $Sol$ are
all homeomorphic to $\mathbb{R}^3$. $S^3$ and $S^2 \times R^1$
provide the other two topologies. (Koike \textit{et al} 1994,
p19).

Note that not all of the globally homogeneous, simply connected
$3$-dimensional Lie groups from the Bianchi classification admit a
compact quotient. For example, Bianchi type $\textrm{IV}$ and the
one-parameter family in Bianchi type $\textrm{VI}_h$ do not admit
a compact quotient, and therefore do not provide a Thurston
geometry.

\hfill \break

Assuming the Copernican principle is true, the observed local
isotropy of our universe can be used to exclude any $3$-manifold
which is not a prime manifold. A prime manifold is a manifold
which has no non-trivial connected sum decomposition. Primeness is
a necessary condition for a $3$-manifold to accept a metric of
constant sectional curvature, hence any non-trivial connected sum
of prime manifolds can be excluded as a candidate for the spatial
topology of our universe. Note that any compact $3$-manifold can
be decomposed as a finite connected sum of prime $3$-manifolds,
and any compact orientable $3$-manifold can be decomposed as a
\emph{unique} finite connected sum of primes. Although $M \# S^3
\cong M$ for any $3$-manifold $M$, the connected sum construction
provides a method of obtaining a plentiful family of compact
orientable $3$-manifolds which are inconsistent with the
conjunction of the Copernican principle and our observation of
local isotropy. This should be balanced with Thurston's assertion
that `most' compact orientable $3$-manifolds accept a metric of
constant negative curvature.

Note also that whilst primeness is a necessary condition for a
$3$-manifold to accept a metric of constant sectional curvature,
it is not a sufficient condition. $S^2 \times S^1$ is a prime
manifold, but it cannot accept a metric of constant sectional
curvature, as noted above from the fact that its universal cover
is $S^2 \times \mathbb{R}^1$. To reiterate, only a $3$-manifold
with either $\mathbb{R}^3$ or $S^3$ as universal cover can accept
a metric of constant sectional curvature.

\section{The epistemology of cosmology}

To elucidate the nature and scope of astronomical and cosmological
knowledge, the philosophical purpose of this section is to
precisely clarify, using the concept of the celestial sphere, the
relationship between general relativity and astronomical
observation and measurement. \textit{En route}, the nature of
colour in astronomical observation is clarified, and an
iconoclastic scenario suggested by Arp \textit{et al} (1990) is
used as a case study of the relationship between astronomical
observation and cosmological theory. The nature of the Cosmic
Microwave Background Radiation (CMBR), and its variations, is
clarified, together with a definition and explanation of the
angular power spectrum. The paper concludes with some comments on
the overall status of the FRW models.

\hfill \break

The mathematical formalism of general relativity can be connected
to empirical observation and measurement by means of the concept
of the celestial sphere. One can associate a celestial sphere with
each point of each timelike curve in a Lorentzian manifold
$(\mathcal{M},g)$. In general relativity, the history of an
idealised observer is represented by a timelike curve $\gamma:I
\rightarrow \mathcal{M}$ in a Lorentzian manifold
$(\mathcal{M},g)$, which is such that the tangent to the curve at
each point is a future-pointing, timelike unit vector, (Sachs and
Wu 1977, p41). Hence, one can associate a celestial sphere with
each moment in the history of an idealised observer. At each
moment $\tau$ in the proper time of an observer, there is a
corresponding point $p=\gamma(\tau)$ in the manifold. The tangent
to the curve $\gamma$ at $p$, denoted as $Z$, determines a direct
sum decomposition of the tangent space $T_p \mathcal{M}$:

$$
\mathbb{R}Z \oplus Z^\perp \,.
$$ $\mathbb{R}Z$, the span of $Z$, is the local time axis, and
$Z^\perp$, the set of vectors orthogonal to $Z$, represents the
local rest space of the observer. $Z^\perp$ is isometric with
$\mathbb{R}^3$, and the observer's celestial sphere is the sphere
of unit radius in $Z^\perp$. One can consider the pair $(p,Z)$ as
an instantaneous observer, (Sachs and Wu 1977, p43). Each
instantaneous observer has a private celestial sphere.

Recall now that a light ray/photon is represented by a null
geodesic, and the tangent vector at each point of a null geodesic
is the energy-momentum of the photon. The observation of an
incoming light ray/photon by an instantaneous observer $(p,Z)$,
will be determined by the energy-momentum tangent vector $Y$ of
the null geodesic at $p$.

Given a vector space equipped with an indefinite inner product,
$g(\; ,\; )$, and given an orthonormal basis $\{e_1,...,e_n\}$
such that $\epsilon_i = g(e_i,e_i)$, any vector $v$ in the space
can be expressed as

$$
v = \epsilon_1 g(v,e_1)e_1 + \cdots + \epsilon_n g(v,e_n)e_n \,.
$$ Given that $g(Z,Z) = -1$, the direct sum decomposition determined
by $Z$ enables one to express an arbitrary vector $Y \in T_p
\mathcal{M}$ as

$$
Y = -g(Y,Z)Z + P \, ,
$$ where $P$ is a spacelike vector in the local rest space
$Z^\perp$. There is a unit spacelike vector $B$ such that $P =
bB$, for some real number $b$. Letting $e=-g(Y,Z)$, it follows
that an arbitrary vector $Y$ can be expressed as

$$
Y = eZ + bB \,.
$$

In the case of interest here, where $Y$ is the energy-momentum
tangent vector of a null geodesic at $p$, the null condition means
that $\langle Y,Y \rangle = 0$. This entails that

$$\eqalign{
\langle eZ+bB,eZ+bB\rangle &= \langle eZ,eZ \rangle + \langle
bB,bB \rangle \cr &= - e^2 + b^2 \cr &= 0 \, ,}$$ which is
satisfied if and only if $e=b$. Hence, for a null vector $Y$,

$$
Y = eZ +eB \,.
$$ Letting $U = -B$, we have

$$
Y = e(Z-U) = -g(Y,Z)Z + g(Y,Z)U \, ,
$$ with $P=-eU$. $U$ is a unit spacelike vector in the celestial
sphere, pointing in the spacelike direction from which the photon
with the null vector $Y$ emanates.

The instantaneous observer $(p,Z)$ will detect the photon of light
to be of energy $e = -g(Y,Z) \in (0,\infty)$, and to come from the
spatial direction $U \in Z^\perp$, where $Y = e(Z-U)$, (Sachs and
Wu 1977, p46 and p130). The measured frequency of the light will
simply be $\nu=e/h$, and the wavelength will be $\lambda=c/\nu$,
or simply $\lambda=\nu^{-1}$ if `geometric units' are used, in
which $c=1$.

Observers in a different state of motion at the same point $p$ in
space, will be represented by different timelike vectors at $p$.
Two distinct timelike vectors $V,W \in T_p \mathcal{M}$ will
determine different direct sum decompostions of $T_p \mathcal{M}$.
As a consequence, observers in a different state of motion will
have different local rest spaces, $V^\perp$ and $W^\perp$, and
will have different celestial spheres. This results in the
aberration of light: different observers will disagree about the
position of a light source, (Sachs and Wu 1977, p46). Moreover,
different observers at a point $p$ will measure the same photon of
light to have different energy. $(p,V)$ would measure $e =
-g(Y,V)$, and $(p,W)$ would measure $e = -g(Y,W)$.

In the simplest of cases, the colour of an object perceived by an
observer is determined by the energy, within the visible spectrum,
at which most of the photons are emitted or reflected from that
object. Hence, the colour of light detected from some source will
be dependent upon one's motion with respect to the source. Let us
agree to define an intrinsic property of an object to be a
property which the object possesses independently of its
relationships to other objects. Let us also agree to define an
extrinsic property of an object to be a property which the object
possesses depending upon its relationships with other objects. The
colour of an object, determined by the energy of the light it
emits or reflects, is not an intrinsic property of an object. The
colour of an object varies depending upon the relationship between
that object and an observer, hence the colour of an object is an
extrinsic property. The perception of colour has a number of
additional subtleties associated with it, which we will detail at
a later juncture.

We are ultimately interested in cosmology, so we shall consider
here only the way in which the formalism of general relativity is
linked with astronomical observations. Given an instantaneous
observer $(p,Z)$, one can associate with it a celestial sphere
$\mathscr{S}_Z$ and a direction-energy space $\mathscr{S}_Z \times
(0,\infty)$, (Sachs and Wu 1977, p141). Recall that each photon,
corresponding to a forward-pointing null vector $Y \in
T_p\mathcal{M}$, has an energy $e = -g(Y,Z)$ and a spatial
direction $U \in \mathscr{S}_Z$, hence the notion of a
direction-energy space. $\mathscr{S}_Z \times (0,\infty)$ is
diffeomorphic to the forward light cone $\mathscr{V}_0^+$, which
in turn is diffeomorphic to $Z^\perp - \textbf{0} \cong
\mathbb{R}^3 - \textbf{0}$, (Sachs and Wu 1977, p147). Hence, one
can introduce spherical coordinates $(e,\theta,\phi)$ in which the
radial coordinate corresponds to the energy $e$. In these
coordinates, the Euclidean metric tensor on $Z^\perp \cong
\mathbb{R}^3$ can be expressed as

$$
g = e^2(d\theta^2 + \sin^2 \theta d\phi^2) \,.
$$ In these coordinates, the determinant of the metric is
$\det g = e^4 \sin^2 \theta$. The natural metric volume element of
a Riemannian metric $g$ in a coordinate system $(x_1,...,x_n)$ is
defined to be

$$
\Omega = (\sqrt{| \det g | }) \, dx^1 \wedge \cdots \wedge dx^n \,
, $$ hence in the case above, the natural metric volume element is

$$\eqalign{
\Omega &= e^2 \sin \theta \; de \wedge d\theta \wedge d\phi \cr &=
de \wedge e d\theta \wedge e \sin \theta d\phi \cr &= de \wedge
e^2(d\theta \wedge \sin \theta d\phi) \cr &= de \wedge e^2\omega
\, ,}
$$ where $\omega$ is the standard metric volume element on the
$2$-sphere.

Sachs and Wu introduce a photon distribution function $N_Z$ on the
direction-energy space of an instantaneous observer, (1977, p142),

$$N_Z:\mathscr{S}_Z \times (0,\infty) \rightarrow [0,\infty) \,.$$

Given that $de \wedge e^2\omega = e^2de \wedge \omega$, for a
range of energies $[a,b] \subset (0,\infty)$ and a compact subset
of the celestial sphere $\mathscr{K} \subset \mathscr{S}_Z$,

$$
\int_{\mathscr{K} \times [a,b]}N_Z \, \Omega = \int_\mathscr{K}
\omega \int_a^b e^2N_Z \, de \,.
$$

Sachs and Wu define this integral to be the number of photons per
unit spatial volume in the energy range $[a,b]$ emanating from the
compact region $\mathscr{K}$ of the observer's celestial sphere,
(p142). They interpret $e^2N_Z$ as the number of photons per unit
spatial volume per unit solid angle per unit energy interval. As
they subsequently explain, (p147-148), because photons travel at
unit speed in the `geometric units' employed, they travel a unit
distance in unit time. Hence, the number of photons which occupy a
unit spatial volume is equal to the number of photons which pass
through a unit area perpendicular to their direction of motion in
unit time. Therefore $e^2N_Z$ can also be interpreted as the
number of photons which pass through a unit perpendicular area per
unit time per unit solid angle upon the celestial sphere per unit
energy interval. In terms of astronomical observations, the unit
area is the unit area of some photon collection device, such as
the surface of a radio telescope, or the mirrored surface of an
optical telescope.

Making the independent variables explicit, $e^2N_Z$ is a function
$e^2N_Z(a,t,\theta,\phi,e)$, where $a$ denotes a point on the
surface on the photon collection device, $t$ denotes time,
$\theta$ and $\phi$ are coordinates upon the celestial sphere of
the instantaneous observer, and $e$ is the energy. A different
function $e^3N_Z(a,t,\theta,\phi,e)$ specifies the amount of
energy passing through a unit perpendicular area per unit time per
unit solid angle upon the celestial sphere per unit energy
interval. When $e$ is replaced with the frequency of the
radiation, $\nu = e/h$, the function $e^3N_Z(a,t,\theta,\phi,\nu)$
specifies the amount of energy passing through a unit
perpendicular area per unit time per unit solid angle upon the
celestial sphere per unit frequency interval. In the astronomy
literature, this function is referred to as the \emph{specific
intensity} of radiation. Its dimensions are Watts ($W$) per square
metre ($m^{-2}$) per Hertz ($Hz^{-1}$) per steradian
($sterad^{-1}$). The specific intensity is often denoted as
$I_\nu$ to emphasise that it is a function of the frequency $\nu$
of radiation. In this event, $I$ is often reserved to denote the
integral of the specific intensity over all frequencies

$$
I = \int_0^\infty I_\nu \, d\nu \,.
$$ The resulting function $I(a,t,\theta,\phi)$ specifies the amount
of energy passing through a unit perpendicular area per unit time
per unit solid angle upon the celestial sphere, over all
frequencies.

Suppose that a light source such as a star, a nebula or a galaxy
corresponds to a compact region $\mathscr{K}$ upon the celestial
sphere of an observer. The \emph{flux density} $F$ of the light
source is obtained by integrating the intensity $I$ over the
region $\mathscr{K}$. To be precise, one integrates $I \cos \alpha
$, where $\alpha$ is the angle between each point in $\mathscr{K}$
and the perpendicular to the surface area of the measuring device,
(Karttunen \textit{et al} 2003, p81). In the case of a light
source which subtends a small solid angle upon the celestial
sphere, and a measuring instrument pointed directly at the light
source, $\cos \alpha \approx 1$. One can deal with either a
frequency-dependent flux
$$F_\nu = \int_{\mathscr{K}}I_\nu \cos \alpha \; \omega \, ,$$ or the total flux $$F=
\int_\mathscr{K} \omega \int_0^\infty I_\nu \cos \alpha \; d\nu
\,.
$$ The dimensions of $F_\nu$ are $W \, m^{-2} \, Hz^{-1}$, whilst
the dimensions of $F$ are $W \, m^{-2}$.

The flux density $F(r)$ observed from a light source at a distance
$r$ is another name for the \emph{apparent luminosity} $l(r)$ of
the light source at distance $r$. Assuming space is approximately
Euclidean on the length scales involved, and assuming that the
light is emitted isotropically from the source, the \emph{absolute
luminosity} $L$ of the light source is defined to be $L = 4\pi r^2
F(r)$. The absolute luminosity is simply the power output of the
light source, the amount of energy emitted per unit time, in all
directions. That power is spread out over spheres of increasing
surface area $4\pi r^2$ at increasing distances $r$, hence the
flux decreases as a function of distance $F(r) = L/4\pi r^2$.

The brightness of an object, either in astronomy, or in perception
with the naked eye, corresponds not to the specific intensity of
the light received from that object, but to the flux density of
the light. Assuming that an object and observer are not in
relative motion and that the space between the object and observer
is static, then the specific intensity of the light received from
the object is independent of the distance separating the observer
from the object, whilst the flux density is inversely proportional
to the square of the distance, (Karttunen \textit{et al} 2003,
p89). If an object and observer are either in relative motion, or
the space between them is dynamic, then the flux density will also
depend upon the redshift/blueshift.

Sachs and Wu suggest (p142) that the brightness of a rose
corresponds to the specific intensity $e^3N_Z$. The specific
intensity is independent of distance because it measures the flux
density per unit solid angle. At greater distances, a unit solid
angle collects photons emitted from a larger fraction of the
surface area of the object, but due to the greater distance, the
unit solid angle collects a smaller fraction of the photons
emitted from the surface area under its purview. These effects
cancel. The brightness of an object to the naked eye decreases
with distance, hence specific intensity does not correspond to the
naked eye perception of brightness.

The brightness of an object to the naked eye corresponds not to
the total flux density of the object, but to the flux integrated
over the visible range of frequencies:

$$\eqalign{
F_{[a,b]} &= \int_\mathscr{K} \omega \int_a^b I_\nu \cos \alpha \;
d\nu \cr &= \int_a^b F_\nu \, d\nu \,.} $$

If the angle subtended by a luminous object remains constant, but
the intensity of the light it radiates increases, then the flux
density $\equiv$ brightness of the object will increase. Hence,
although the brightness of an object should not be conflated with
the intensity of the light radiated by the object, it is
legitimate to explain an increase in the brightness of an object
as being the result of an increase in the intensity of the light
it emits.

Picking up an issue alluded to above, the colour of an object
perceived by an observer is determined by the intensity of the
light emitted or reflected from that object, over the range of
visible wavelengths, in the reference frame of that observer. The
visible spectrum contains those colours which can be identified in
a rainbow, or in the light refracted from a prism. These `spectral
colours' each correspond to a particular wavelength or range of
wavelengths. If the intensity of light over the visible spectrum
is peaked at a certain wavelength in an observer's reference
frame, then that observer perceives the corresponding colour.
However, the human perceptual system introduces colours and
structures amongst the set of colours, which do not exist in the
visible spectrum itself, (Clark 1998). For a start, whilst the
visible spectrum has the topology of a closed interval $[0,1]$ of
the real line, and a consequent linear ordering, the set of
colours perceived by humans has the topology of the circle $S^1$,
and, obviously, no such linear ordering relationship. The visible
spectrum ranges from the blue end at 400nm to the red end at
700nm. A type of purple, called magenta, exists between blue and
red in the set of humanly perceived colours, and completes the
circle.

Magenta can be defined as a mixture of red and blue, and this
introduces the second difference between the visible spectrum and
the set of humanly perceived colours. Let us adopt the common
nomenclature, and refer to the latter as the set of `hues'. One
can mix hues that do correspond to spectral colours, to produce
new hues which don't correspond to spectral colours. Such hues
correspond to an intensity curve which has multiple peaks over the
visible spectrum. Different combinations of hue can produce the
same mixed hue; these hue combinations are called `metamers'. This
means that different intensity curves over the visible spectrum,
with different combinations of wavelength peaks, can produce the
same perception of colour. There is a many-one correspondence
between intensity curves and perceived colours.

In general, three parameters are used to characterise the space of
colours in the human perceptual system. The exact parameters used
depend upon whether one is dealing with reflected light from a
surface, emitted light from a source, or the light which falls
upon a photographic emulsion after passing through an aperture.
With this qualification, the three parameters are hue, saturation,
and lightness, the latter sometimes being thought of as the shade
of a colour. Shade is the relative amount of lightness or darkness
of a colour. For a particular hue, you get a lighter shade by
mixing it with white, and a darker shade by mixing it with black.
Lightness measures the overall intensity of a colour; lighter
shades are therefore brighter. The saturation of a colour measures
the ratio of the intensity at the dominant wavelengths to the
intensity at other wavelengths. If the dominant wavelengths of a
hue are highly peaked, then that hue has high saturation. If the
peaks are quite small compared to the intensity at other
wavelengths, then the hue tends towards an achromatic grey, and is
said to have low saturation. Pastel colours are low saturation
hues. For achromatic light, the lightness scale ranges from white
to black through all the various intervening greys. No light at
all at visible wavelengths produces the perception of black. Equal
combinations of light at different wavelengths within the visible
spectrum produce achromatic light, and each hue has a complement,
such that when that hue is combined with its complement, the
result is achromatic light.

One can treat hue, saturation, and lightness as cylindrical polar
coordinates upon the space of colours in the human perceptual
system. The circle of hues has the angular coordinate, saturation
provides a radial coordinate in the plane, and lightness provides
the `vertical' coordinate. Note, however, that for darker shades,
the saturation range is more restricted, so one is dealing with
something more akin to a cone than a cylinder. Note also that
there are other coordinatizations in use, such as the `colour
sphere'.

At a rather high level of idealisation, Sachs and Wu (1977, p142)
suggest that one can regard all astronomers who have ever lived as
a single instantaneous observer $(p,Z)$. I will slightly relax
this idealisation, and suggest instead that one can associate a
single celestial sphere with the human race. Whilst each
individual has a private celestial sphere, at another level of
idealisation there is a celestial sphere which is common to all
humans upon the Earth. Gazing skywards on a clear night, the stars
appear to be speckled across the inner surface of an inverted
bowl. This is one hemisphere of our common celestial sphere. The
history of the human race can be represented as a timelike curve,
and as Sachs and Wu suggest (1977, p131), one can use parallel
transport to identify the celestial spheres associated with the
points of a timelike curve. Thence, (changing notation slightly),
all the possible astronomical observations made by the human race
could be encoded as a time-dependent function of only three
variables $I_t(\theta,\phi,\nu)$. The function $I_t$ specifies the
intensity at time $t$ of electromagnetic radiation at any
frequency $\nu$ over the entire celestial sphere. The time
variation of this function provides all the raw astronomical data
that a species located upon a single planet could ever have. In
terms of using the raw data upon our celestial sphere to make
cosmological inferences, it should be noted that only $1\%$ of the
light which intersects our celestial sphere comes from beyond our
galaxy, (Disney 2000, p4).

The conventional coordinates upon a sphere are such that $\theta
\in [0,\pi]$ and $\phi \in [0,2\pi)$. Astronomers use a variety of
slightly different, but closely related celestial coordinates. For
example, the \emph{equatorial system} (Nicolson 1977, p42-43) uses
the intersection of the plane of the Earth's equator with the
celestial sphere to determine a great circle on the celestial
sphere called the celestial equator. Right ascension $\alpha \in
[0,2\pi)$ then provides a coordinate upon the celestial equator,
starting at the vernal equinox and running Eastward. Declination
$\delta \in [-\frac{1}{2}\pi,\frac{1}{2}\pi]$ then specifies the
angular distance North or South of the celestial
equator.\footnote{The vernal equinox is the point of intersection
of the ecliptic and the celestial equator at which the Sun moves
from the Southern celestial hemisphere into the Northern celestial
hemisphere (Nicolson 1977, p234). The ecliptic is the great circle
which the Sun traces upon the celestial sphere due to the Earth's
annual orbit around the Sun (1977, p73). It can also be thought of
as the intersection of the Earth's orbital plane with the
celestial sphere. Because the Earth's axis, and therefore its
equator, are inclined at approximately $23\frac{1}{2}\deg$ to the
orbital plane, the celestial equator is inclined at the same angle
to the ecliptic. The ecliptic intersects the celestial equator at
two points, the vernal equinox and the autumnal equinox.}

The timelike vector $Z$ that specifies which local rest space, and
thence which celestial sphere, is selected for the human race,
will be determined by taking the vector sum of the motion of the
Local Group of galaxies relative to the microwave background
radiation, the motion of the Milky Way within the Local Group, the
motion of the Sun within the Milky Way, and the motion of the
Earth around the Sun.

General relativity enables us to interpret the complete array of
astronomical images upon the celestial sphere, as the projection
onto the celestial sphere of all the light sources contained
within our past light cone $E^-(x)$. The past light cone $E^-(x)$
of our point in space-time $x \in \mathcal{M}$, is a
$3$-dimensional null hypersurface whose universal covering is a
manifold of topology $S^2 \times \mathbb{R}^1$. One can use the
right ascension and declination coordinates $(\alpha,\delta)$ upon
the $S^2$ factor, and in a simple type of expanding universe, one
can use redshift $z$ as the $\mathbb{R}^1$-coordinate.

To interpret the raw data $I_t(\theta,\phi,\nu)$ upon the
celestial sphere it is necessary to use theories of light emission
and absorption processes. These theories enable us to interpret
the raw data in terms of the electromagnetic spectra of chemical
elements and compounds, and in terms of the statistical mechanics
and thermodynamics of the matter which either emits the radiation,
or absorbs some parts of it.

The best example of this is provided by the cosmic microwave
background radiation (CMBR). This radiation has a spectrum which
is very close to that of `Planckian' blackbody radiation, often
called thermal radiation. Blackbody radiation at temperature $T$
is radiation whose specific intensity is given by, (Sachs and Wu,
p144-145),

$$
I_\nu = e^3N_Z = e^3(2h^{-3}[exp(h\nu/kT)-1]^{-1}) \, ,
$$ where $k$ is the Boltzmann constant.

It is known both from theory, and from Earth-bound experiment and
observation, that only radiation which is in a state of
equilibrium with matter can have a blackbody spectrum. The
radiation is said to be `thermalised' by its interaction with
matter. It is only when there is no net transfer of energy between
the radiation and the matter, that the radiation will be
blackbody. Deep inside a star, where the gas is opaque, the
radiation will be blackbody radiation. Similarly, the radiation
inside the evacuated cavity of an opaque-walled box, whose walls
are maintained at a constant temperature, will be blackbody. The
opacity is necessary because it is the interaction between the
matter and the radiation which makes the radiation blackbody.

Now, as Layzer puts it ``the present day Universe is just as
transparent to the [microwave] background radiation as it is to
ordinary light. We are not living in the equivalent of an opaque
box or inside an opaque gas. This means that the background [i.e.
the CMBR] could not have acquired its distinctive blackbody
characteristics under present conditions. The background radiation
must be a relic of an earlier period of cosmic history, when the
Universe was far denser and more opaque," (Layzer, 1990, p147).

Although the reasoning here is correct, Arp \textit{et al} (1990)
challenged the empirical claim that the present universe is
effectively transparent to radiation at all wavelengths. It is
commonly believed that radiation emitted from stars is able to
propagate freely through space, with only negligible absorption
and scattering by interstellar/intergalactic gas and dust, and
planets. The matter which does absorb radiation is distributed in
a clumpy, discrete manner across the sky, yet the CMBR is
continuum radiation across the entire celestial sphere. Thus, it
is reasoned, the CMBR could not have been produced in the present
universe.

Arp \textit{et al} argued that the CMBR we observe, was emitted
recently and locally. They suggested that there is some form of
intergalactic material, ``with the property of being strongly
absorptive of microwaves, yet of being almost translucent in both
the visible and longer radio wave regions of the spectrum," (1990,
p809). They suggested that our present universe is opaque in the
microwave, and that starlight is absorbed and scattered by this
intergalactic material to produce an isotropic blackbody microwave
spectrum across our celestial sphere. Although stars are discrete
sources of light, because the hypothetical intergalactic material
is distributed uniformly, it could produce a continuum of
radiation across the celestial sphere.

When a photon of starlight is absorbed by interstellar gas, the
gas re-radiates the energy that is absorbed, but it does so by
emitting a sequence of lower-energy photons, and it emits the
photons in random directions. This characteristic might be able to
explain the isotropy of the CMBR. The intergalactic material might
be re-radiating starlight equally in all directions.

If the present universe were opaque in the microwave, it would no
longer follow that the CMBR must be a relic of an earlier period
of cosmic history. One of the primary pillars providing empirical
support for FRW cosmology would have crumbled.

Arp \textit{et al} suggested that metallic filaments, particularly
iron filaments, blasted into intergalactic space by supernovae,
would provide the requisite microwave opacity. Arp \textit{et al}
concluded quite splendidly ``The commonsense inference from the
planckian nature of the spectrum of the microwave background and
from the smoothness [i.e.uniformity] of the background is that, so
far as microwaves are concerned, we are living in a fog and that
the fog is relatively local. A man who falls asleep on the top of
a mountain and who wakes in a fog does not think he is looking at
the origin of the Universe. He thinks he is in a fog," (1990,
p810).

At the risk of sounding churlish, the case of a man who falls
asleep atop a mountain is not relevantly analogous to the
astronomical predicament of the human race. If we had made
observations of distant objects in the microwave for some years,
without any impediment, but after a period of academic sleep, we
then returned to find an isotropic obscuration in the microwave,
we would indeed be justified in thinking that a microwave fog had
developed. The position of the human race is that we have found a
microwave fog from the time that we began looking.

It is well-known that Alpher and Herman predicted in 1948, using
FRW cosmology, that the present universe should be permeated by a
residue of electromagnetic radiation from the early universe. This
radiation was detected by Penzias and Wilson in 1965. Rhook and
Zangari point out that ``because the existence of a background of
microwave radiation was predicted as a consequence of the big
bang, its account, unlike that of rivals, was granted immunity
against accusations of being ad hoc. Competing theories were then
forced into constructing post hoc explanations for the radiation
which did not carry the force of being prior predictions, and
which themselves lay open to charges of being ad hoc," (1994,
p230).

According to a FRW model of our universe there was no net transfer
of energy between the radiative component of the energy density
and the matter component of the energy density, until the universe
was $10^4-10^5$ yrs old. At that time, the `epoch of last
scattering', the universe had expanded to the point that the
equilibrium reactions between the photons and the plasma of matter
could no longer be maintained, and the universe became transparent
to all but a negligible fraction of the radiation. Blackbody
radiation was emitted throughout space, and the FRW models
represent this radiation to cool as the universe expands, until it
reaches microwave frequencies in the present era. The FRW models
therefore predict the continuum, blackbody, microwave radiation
that we observe today.

The verification of FRW cosmology by the detection of the CMBR is
the hypothetico-deductive method at its finest. The physical
processes responsible for the CMBR cannot be deduced from the
empirical characteristics of the CMBR, as the work of Arp
\textit{et al} demonstrates. Instead, one hypothesizes the FRW
models, one deduces the empirical predictions, and one compares
and verifies the predictions with the astronomical data. The mere
possibility that there could be an alternative explanation for the
CMBR, is not a decisive argument against FRW cosmology.

\hfill \break

The CMBR observed by the COBE and WMAP satellites, and a variety
of Earth-bound/balloon-borne measuring devices, possesses an
approximately blackbody spectrum across the entire celestial
sphere, for all values of $\theta$ and $\phi$. However, the
temperature of the blackbody spectrum varies as a function of
$\theta$ and $\phi$. The CMBR has a blackbody spectrum in all
directions, but there are different blackbody curves in different
directions. The temperature $T$ of the CMBR is a real-valued
function $T(\theta,\phi)$ upon the celestial sphere. Let $\langle
T \rangle$ denote the mean temperature, averaged over the entire
celestial sphere. The function

$$
\delta T(\theta,\phi) = \frac{T(\theta,\phi)- \langle T
\rangle}{\langle T \rangle} \equiv \frac{\Delta T}{\langle T
\rangle}(\theta,\phi)
$$ expresses the temperature deviations (or `fluctuations') as a
fraction of the mean temperature (Coles and Lucchin 1995, p92).
This temperature fluctuation function is itself a real-valued
function upon the celestial sphere, and one can decompose it into
an infinite linear combination of the spherical harmonic functions
upon the sphere, (Coles and Lucchin 1995, p366),

$$
\delta T(\theta,\phi) = \sum_{l = 0}^{\infty} \sum_{m = -l}^{m=+l}
c_{lm} Y^m_l(\theta,\phi) \,.
$$ Note that on a specific celestial sphere, the coefficients
$c_{lm}$ which define the function $\delta T(\theta,\phi)$ are not
functions of $(\theta,\phi)$ themselves. $\delta T(\theta,\phi)$
is a function of $(\theta,\phi)$ because the $Y^m_l(\theta,\phi)$
are functions of $(\theta,\phi)$. The coefficients $c_{lm}$ only
vary across the statistical ensemble of all possible celestial
spheres within our universe.

The spherical harmonics $\{Y^m_l(\theta,\phi): \, l \in
\mathbb{N}, \, m \in (-l,-l+1,...,+l)\}$ form an orthonormal basis
of the Hilbert space $L^2(S^2)$ of square-integrable functions
upon the sphere. They can be defined as

$$
Y^m_l(\theta,\phi) = N^m_l P^{| m |}_l(\cos \theta)e^{im\phi} \, ,
$$ with $ N^m_l$ a normalization constant, and $P^{| m
|}_l(u)$ a Legendre function. Any square-integrable function
$f(\theta,\phi)$ on $S^2$ can then be expressed as a linear
combination

$$
f(\theta,\phi) = \sum_{l = 0}^{\infty} \sum_{m = -l}^{m=+l} c_{lm}
Y^m_l(\theta,\phi) \, ,
$$ with the spherical harmonic coefficients $c_{lm}$ given by

$$
c_{lm} = \langle Y^m_l,f \rangle = \int_{S^2}
\overline{Y}^m_l(\theta,\phi)f(\theta,\phi) \, d\Omega \,.
$$ Note that the angular brackets here denote the inner product on
the space of functions on $S^2$, not to be confused with the use
of angular brackets to denote a mean value.

Physicists tend to refer to the terms in a spherical harmonic
decomposition as `modes'. The term corresponding to $l=0$ is
referred to as the monopole term, $l=1$ terms are called dipole
terms, $l=2$ terms are quadrupole terms, etc. A dipole anisotropy
in the temperature of the CMBR is a periodic variation which
completes $1$ cycle around the sky; it has one `hot' pole and one
`cold' pole. A quadrupole anisotropy is a periodic variation in
the temperature of the CMBR which completes $2$ cycles around the
sky. Mode $l$ anisotropies complete $l$ cycles around the sky.
Higher $l$ modes correspond to temperature fluctuations on smaller
angular scales. For higher $l$ modes, the angular scale
$\vartheta$ of the fluctuation is $\vartheta \approx 60\deg/l$,
(Coles and Lucchin 1995, p367). After subtracting the effects of
the Earth's diurnal rotation, its orbit around the Sun, the motion
of the Sun within the Milky Way galaxy, and the motion of the
Milky Way within the Local Group, we observe from the Earth a
dipole anisotropy in the CMBR upon the celestial sphere. This is a
dipole anisotropy upon our own private celestial sphere due to the
proper motion of the Local Group of galaxies at approx. $600km
s^{-1}$. This dipole anisotropy in the temperature of the CMBR can
be expressed as (Coles and Lucchin 1995, p93)

$$
T(\vartheta) = \langle T \rangle + \Delta T_{dipole} \cos
\vartheta \,.
$$

It is only when one calculates the effect of the proper motion of
the Local Group, and one `subtracts' that effect from the observed
CMBR, that one obtains radiation which is uniform across the
celestial sphere, to at least one part in $10,000$, $\Delta
T/\langle T \rangle < 10^{-4}$, on any angular scale. After
compensating for the effect of our proper motion, the average
temperature of the CMBR is approximately $2.7K$.

The COBE satellite discovered in 1992 that superimposed upon the
dipole temperature anisotropy, there are very small scale
variations in the temperature of the microwave blackbody spectrum
across the entire celestial sphere.

Because radiation was in equilibrium with matter just before they
decoupled, the variations in the CMBR indicate variations in the
density of matter at the time of decoupling. These variations are
believed to be the origins of what have today become galaxies. In
a FRW model, the subsequent formation of galaxies has a negligible
effect upon the CMBR. Thus, the variations in the CMBR are thought
to indicate inhomogeneity at the so-called `epoch of last
scattering'.

Of deep observational significance at the present time is the CMBR
\emph{angular power spectrum}. To clarify precisely what this is,
it will be necessary to carefully distinguish between two
different mathematical expressions. To obtain the first
expression, begin by noting that whilst the mean value of the
temperature fluctuations is zero, $\langle \delta T \rangle = 0$,
the variance, the mean value of the square of the fluctuations,
$\langle (\delta T)^2 \rangle$, is non-zero.

Consider $| \delta T |^2(\theta,\phi)$. Given the expansion of
$\delta T$ in the spherical harmonics, it follows that

$$
| \delta T |^2(\theta,\phi) = \sum_{l = 0}^{\infty} \sum_{m =
-l}^{m=+l} \sum_{l' = 0}^{\infty} \sum_{m' = -l'}^{m'=+l'}
c_{lm}^* c_{l'm'}
\overline{Y}^m_l(\theta,\phi)Y^{m'}_{l'}(\theta,\phi) \,.
$$ $\overline{Y}^m_l(\theta,\phi)$ and $Y^{m'}_{l'}(\theta,\phi)$
don't vary over the ensemble of all celestial spheres, so if
$\langle | \delta T |^2 \rangle (\theta,\phi)$ is taken to be the
mean value of $| \delta T |^2(\theta,\phi)$ over the ensemble, it
can be expressed as

$$
\langle | \delta T |^2 \rangle (\theta,\phi) = \sum_{l =
0}^{\infty} \sum_{m = -l}^{m=+l} \sum_{l' = 0}^{\infty} \sum_{m' =
-l'}^{m'=+l'} \langle c_{lm}^* c_{l'm'} \rangle
\overline{Y}^m_l(\theta,\phi)Y^{m'}_{l'}(\theta,\phi) \,.
$$ Now, given that $\langle c_{lm}^* c_{l'm'} \rangle = \langle | c_{lm} |^2
\rangle \delta_{l l'}\delta_{m m'}$, this expression reduces to

$$
\langle | \delta T |^2 \rangle (\theta,\phi) = \sum_{l =
0}^{\infty} \sum_{m = -l}^{m=+l} \langle  | c_{lm} |^2 \rangle |
Y^{m}_{l}(\theta,\phi) | ^2 \,.
$$ Noting that $\delta T$ is real-valued, this means

$$
\langle (\delta T) ^2 \rangle (\theta,\phi) = \sum_{l =
0}^{\infty} \sum_{m = -l}^{m=+l} \langle  | c_{lm} |^2 \rangle |
Y^{m}_{l}(\theta,\phi) | ^2 \,.
$$

This expression is clearly dependent on $(\theta,\phi)$. A second
approach yields an expression with no such dependence:

The function $\delta T(\theta,\phi)$ is a vector in the Hilbert
space of functions $L^2(S^2)$. This space of functions, as a
Hilbert space, is equipped with an inner product $\langle \; , \;
\rangle$, and a norm $\| \; \|$. (Again, the angular brackets of
the inner product here should not be confused with the angular
brackets that define a mean value). The norm defines the length of
a vector in the vector space of functions. Consider the square of
the norm $\| \delta T \|^2$ of the function $\delta
T(\theta,\phi)$:

$$\eqalign{
\| \delta T \|^2 &= \langle \delta T, \delta T \rangle \cr &=
\left \langle \sum_{l = 0}^{\infty} \sum_{m = -l}^{m=+l} c_{lm}
Y^m_l(\theta,\phi), \sum_{l = 0}^{\infty} \sum_{m = -l}^{m=+l}
c_{lm} Y^m_l(\theta,\phi) \right \rangle \cr &= \sum_{l =
0}^{\infty} \sum_{m = -l}^{m=+l} | c_{lm} |^2 \,.}
$$ This follows because $\langle Y^m_l(\theta,\phi),
Y^{m'}_{l'}(\theta,\phi)\rangle = \delta_{l l'}\delta_{m m'}$

Using angular brackets to denote the mean once again, $\langle \|
\delta T \|^2 \rangle$ denotes the mean of the squared length of
the function vector, taken over all possible celestial spheres.
From the last expression, it follows that

$$
\langle \| \delta T \|^2 \rangle = \sum_{l = 0}^{\infty} \sum_{m =
-l}^{m=+l} \langle | c_{lm} |^2 \rangle \,.
$$ This is the sum of the mean of the square modulus value of all the
coefficients from the spherical harmonic expansion of $\delta T$.
The mean $\langle | c_{lm} |^2 \rangle$ is the mean of $| c_{lm}
|^2$ taken over the ensemble of celestial spheres. $| c_{lm} |^2$
is fixed for each celestial sphere.

By an ergodic hypothesis, for large $l$ this average is
approximated by an average taken over all the modes with the same
$l$ on our private celestial sphere

$$
\sum_{m = -l}^{m=+l} \langle | c_{lm} |^2 \rangle =
\frac{1}{(2l+1)}\sum_{m=-l}^{m=+1} | c_{lm} |^2 \,.
$$

The angular power spectrum is

$$
C_l = \frac{1}{(2l+1)}\sum_{m=-l}^{m=+1} | c_{lm} |^2 \,.
$$ Hence

$$
\langle \| \delta T \|^2 \rangle = A + \sum_{l = l_b}^{\infty}C_l
\,.
$$ $A$ is the contribution from the small $l$ spherical harmonics, and
$l_b$ is the lower bound at which the ergodic hypothesis becomes
valid. For large $l$, $C_l$ is the contribution to the mean of the
squared length of the temperature fluctuation function vector from
the mode $l$ spherical harmonics.

The value of $l$ for the highest peak in the CMBR power spectrum
corresponds to hot and cold spots of a specific angular size on
the celestial sphere, (Tegmark 2002, p2). The exact angular size
of these spots can be used to determine if the curvature of space
is positive, negative or zero. If the peak in the CMBR power
spectrum corresponds to spots which subtend a specific value close
to $0.5 \deg$, then space is flat, (2002, p2). If space has
positive curvature, then the angles of a triangle add up to more
than $180 \deg$, and the size of the CMBR spots would be greater
than $0.5 \deg$. If space has negative curvature, then the angles
of a triangle add up to less than $180 \deg$, and the size of the
CMBR spots would be less than $0.5 \deg$. The current data on the
CMBR indicates that the spot size is very close to $0.5 \deg$, but
cannot determine the exact value. Thus, the current data merely
confirms the long-held belief that the curvature of space is very
close to zero.

Tegmark falsely states that ``many of the most mathematically
elegant models, negatively curved yet compact spaces, have been
abandoned after the recent evidence for spatial flatness," (2002,
p3). Unless Tegmark means that the evidence indicates a $k=0$
universe, (which it doesn't), this remark might betray the
misunderstanding of the rigidity theorem for hyperbolic manifolds
alluded to before. Negative values of spatial curvature very close
to zero exclude the possibility of a compact hyperbolic universe
which is sufficiently small for the topology to be detectable, but
it does not exclude the possibility that the spatial universe does
have a compact hyperbolic topology.

The CMBR power spectrum can also be used to determine whether our
spatial universe is a \emph{small} compact universe. Whilst a
compact universe of volume much greater than the Hubble volume
would leave no imprint upon the CMBR, a small compact universe
would affect the CMBR power spectrum on large angular scales, and
could leave paired circles in the CMBR at antipodal positions on
the celestial sphere. No paired circles have been discovered, but
the WMAP satellite has revealed anomalies in the CMBR power
spectrum on large angular scales. The quadrupole $l=2$ mode was
found to be about $1/7$ the strength predicted for an infinite
flat universe, while the octopole $l=3$ mode was $72\%$ of the
strength predicted for such a non-compact $k=0$ universe, (Luminet
\textit{et al} 2003, p3).

Tegmark states that ``the interim conclusion about the overall
shape of space is thus `back to basics': although mathematicians
have discovered a wealth of complicated manifolds to choose from
and both positive and negative curvature would have been allowed
\emph{a priori}, all available data so far is consistent with the
simplest possible space, the infinite flat Euclidean space that we
learned about in high school," (2002, p3). As emphasised above, it
is also the case that all the data remains consistent with
positive or negative curvature, and with multiply connected
topology as well as simply connected topology. No such `back to
basics' conclusion can be drawn.

\hfill \break

The present universe only approximates a FRW model on length
scales greater than $100$Mpc. On smaller length scales, the
universe exhibits large inhomogeneities and anisotropies. The
distribution of matter is characterised by walls, filaments and
voids up to $100$Mpc, with large peculiar velocities relative to
the rest frame defined by the CMBR.

Whilst the CMBR indicates that the matter in the universe was
spatially homogeneous to a high degree when the universe was
$10^{4}-10^{5}$yrs old, the distribution of galaxies is an
indicator of the distribution of matter in the present era, when
the universe is $\sim 10^{10}$ yrs old. Given perturbations from
exact homogeneity which were sufficiently large relative to the
speed of expansion when the universe was $10^{4}-10^{5}$yrs in
age, one would expect the degree of homogeneity to decrease with
the passage of time. Small initial inhomogeneities result in some
regions which are denser than the average. A positive feedback
process then ensues. The regions of greater than average density
gravitationally attract matter from the surroundings, thus
increasing the excess density of matter. As the excess density of
matter increases, a greater force is exerted on the surrounding
matter, thus continuing to increase the agglomeration of matter.
Gravity magnifies small initial inhomogeneities. Hence, the FRW
models become increasingly inaccurate as the universe gets older.
The length scale on which the universe can be idealised as being
homogeneous, grows as a function of time, hence the length scale
on which a FRW model is valid, grows as a function of time. Not
only do the FRW models constitute a first approximation, but they
constitute an increasingly inaccurate first approximation.


\begin{thebibliography}{99}
\bibitem{Arp90}
Arp, H.C., Burbidge, G., Hoyle, F., Narlikar, J.V.,
Wickramasinghe,N.C. (1990). \emph{The extragalactic universe: an
alternative view}, Nature Volume 346, p807-812.
\bibitem{BeemEhrlich81}
Beem, J.K. and Ehrlich, P.E. (1981) \emph{Global lorentzian
geometry}, New York: Dekker.
\bibitem{Besse87}
Besse, A.L. (1987). \emph{Einstein manifolds}, Berlin and New
York: Springer-Verlag.
\bibitem{BlauGuth87}
Blau, S.K. and Guth, A.H. (1987). Inflationary Cosmology. In
S.Hawking and W.Israel (eds.), \emph{300 Years of Gravitation},
(pp524-603). Cambridge: Cambridge University Press.
\bibitem{Boothby86}
Boothby, W.M. (1986). \emph{An introduction to differentiable
manifolds and Riemannian geometry}, 2nd Edition, London: Academic
Press.
\bibitem{Clark98}
Clark, A. (1998) Perception:Color. In W.Bechtel and G.Graham
(eds.), \emph{A companion to cognitive science}, (p282-288).
Oxford: Blackwell.
\bibitem{ColesLucchin95}
Coles, P. and Lucchin, F. (1995). \emph{Cosmology: the origin and
evolution of cosmic structure}, Chichester: John Wiley and Sons.
\bibitem{CornishSpergStark98}
Cornish, N.J., Spergel, D.N., Starkman, G.D. (1998). Circles in
the Sky: Finding Topology with the Microwave Background Radiation.
\emph{Classical and Quantum Gravity} 15, pp2657-2670.
arXiv:astro-ph/9801212 v1 22 Jan 1998.
\bibitem{CornishWeeks98}
Cornish, N.J. and Weeks, J.R. (1998). \emph{Measuring the shape of
the universe}, arXiv:astro-ph/9807311 v2 5 Aug 1998
\bibitem{Disney00}
Disney, M.J. (2000). \emph{The case against cosmology},
arXiv:astro-ph/0009020 v1 1 Sep 2000.
\bibitem{Novikov92}
Dubrovin, M., Fomenko, A.T. and Novikov, S.P. (1992). \emph{Modern
geometry: methods and applications}, 2nd ed, New York:
Springer-Verlag.
\bibitem{Ellis71}
Ellis, G.F.R. (1971). Topology and cosmology. \emph{General
Relativity and Gravitation}, vol.2, No.1, p7-21.
\bibitem{Ellis95}
Ellis, G.F.R. (1995). The physics and geometry of the universe:
changing viewpoints. \emph{Quarterly Journal of the Royal
Astronomical Society}, Vol.34, No.3.
\bibitem{FultonHarris91}
Fulton, W. and Harris, J. (1991). \emph{Representation theory},
New York: Springer-Verlag.
\bibitem{JLFriedman91}
Friedman, J.L. (1991). Space-time topology and quantum gravity. In
A.Ashtekar and J.Stachel (eds.), \emph{Conceptual Problems of
Quantum Gravity}, (pp539-569), Boston: Birkhauser.
\bibitem{GuthStein89}
Guth, A. and Steinhardt, P. (1989). The Inflationary Universe. In
P.C.W.Davies (ed.), \emph{The New Physics}, (pp34-60). Cambridge:
Cambridge University Press.
\bibitem{Heller92}
Heller, H. (1992). \emph{Theoretical foundations of cosmology},
Singapore: World Scientific.
\bibitem{Karttunen03}
Karttunnen, H., Kr$\ddot{o}$ger, P., Oja, H., Poutanen, M.,
Donner, K.J. (2003). \emph{Fundamental astronomy}, Fourth Edition,
Berlin-Heidelberg-New York: Springer Verlag.
\bibitem{Koike94}
Koike, T., Tanimoto, M. and Hosoya, A. (1994). \emph{Compact
homogeneous universes}, arXiv:gr-qc/9405052 v2 25 May 1994
\bibitem{KolbTurner90}
Kolb, E.W. and Turner, M.S. (1990). \emph{The early universe},
Redwood City Calif: Addison-Wesley.
\bibitem{Layzer90}
Layzer, D. (1990). \emph{Cosmogenesis: the growth of order in the
universe},  New York-Oxford: Oxford University Press.
\bibitem{Luminet03}
Lachi$\grave{e}$ze-Rey, M. and Luminet, J-P. (2003). \emph{Cosmic
topology}, arXiv:gr-qc/9605010 v2 9 Jan 2003
\bibitem{Luminet99}
Luminet, J-P. and Roukema, B.F. (1999). \emph{Toplogy of the
universe: theory and observation}, arXiv:astro-ph/9901364 v3 19
May 1999.
\bibitem{LuminetOct03}
Luminet, J-P., Weeks, J.R., Riazuelo, A., Lehoucq, R., and Uzan,
J-P. (2003). \emph{Dodecahedral space topology as an explanation
for weak wide-angle temperature correlations in the cosmic
microwave background}, arXiv:astro-ph/0310253 v1 9 Oct 2003
\bibitem{Nicolson77}
Nicolson, I. (1977). \emph{Astronomy: a dictionary of space and
the universe}, Arrow Books.
\bibitem{Nicolson99}
Nicolson, I. (1999). \emph{Unfolding our universe}, Cambridge:
Cambridge University Press.
\bibitem{ONeill83}
O'Neill, B. (1983). \emph{Semi-riemannian geometry: with
applications to relativity}, London: Academic Press.
\bibitem{RainerSchmidt95}
Rainer, M. and Schmidt, H.J. (1995). \emph{Inhomogeneous
cosmological models with homogeneous inner hypersurface geometry},
arXiv:gr-qc/9507013 v1 6 Jul 1995.
\bibitem{RhookZangari94}
Rhook, G. and Zangari, M. (1994). \emph{Should we believe in the
big bang?: a critique of the integrity of modern cosmology}, PSA
1994, Volume 1, pp228-237.
\bibitem{Riess04}
Riess, A.G. \textit{et al} (2004). \emph{Type Ia supernova
discoveries at $z > 1$ from the Hubble space telescope: evidence
for past deceleration and constraints on dark energy evolution},
arXiv:astro-ph/0402512 v1 23 Feb 2004. Astrophysical Journal, 607,
June 2004, pp665-687.
\bibitem{SachsWu77}
Sachs, R.K. and Wu, H. (1977). \emph{General relativity for
mathematicians}, New York and Berlin: Springer-Verlag.
\bibitem{Scott83}
Scott, P.(1983). The geometries of 3-manifolds.
\emph{Bull.London.Math.Soc.} 15, pp401-487.
\bibitem{Sneed71}
Sneed, J.D. (1971). \emph{The Logical Structure of Mathematical
Physics}, Dordrecht: Reidel.
\bibitem{Suppe89}
Suppe, F. (1989). \emph{The Semantic Conception of Theories and
Scientific Realism}, Urbana, Illinois: University of Illinois
Press.
\bibitem{Suppes69}
Suppes, P. (1969). \emph{Studies in the Methodology and Foundation
of Science: Selected Papers from 1951 to 1969}, Dordrecht: Reidel.
\bibitem{Tegmark02}
Tegmark, M. (2002). \emph{Measuring spacetime, from big bang to
black holes}, arXiv:astro-ph/0207199 v1 10 Jul 2002.
\bibitem{ThurstonWeeks94}
Thurston, W.P. and Weeks, J.R.(1984). The mathematics of
three-dimensional manifolds. In \emph{Scientific American}, July
1984, pp94-106.
\bibitem{Wolf67}
Wolf, J.A. (1967). \emph{Spaces of constant curvature}, London:
McGraw-Hill.
\end{thebibliography}
\end{document}